\documentclass[twocolumn, amssymb, aps, showpacs,preprintnumbers,
amsmath,showkeys,floatfix]{revtex4}

\usepackage{graphics}
\usepackage{graphicx}
\usepackage{bm}
\usepackage{epsfig}
\usepackage{times}
\usepackage{subfigure}
\usepackage{color}

\begin{document}

\title{ Quantum interference effects in an ensemble of $^{229}$Th nuclei interacting with coherent light}

\author{Sumanta \surname{Das}}
\email{Sumanta.Das@mpi-hd.mpg.de}

\author{Adriana \surname{P\'alffy}}
\email{Palffy@mpi-hd.mpg.de}

\author{Christoph~H. \surname{Keitel}}
\email{Keitel@mpi-hd.mpg.de}

\affiliation{Max-Planck-Institut f\"ur Kernphysik, Saupfercheckweg 1, D-69117 Heidelberg, Germany}
\date{\today}

\begin{abstract}
As a unique feature, the $^{229}$Th nucleus has an isomeric transition in the vacuum ultraviolet that can be accessed 
by optical lasers. The interference effects occurring in the interaction between coherent optical light and an ensemble of 
$^{229}$Th nuclei are investigated theoretically. We consider the scenario of nuclei doped in vacuum ultraviolet-vacuum 
ultraviolet transparent crystals and take into account the effect of different doping sites and therefore different lattice fields 
that broaden the nuclear transition width. This effect is shown to come in interplay with interference effects due to the hyperfine 
splitting of the ground and isomeric nuclear states. We investigate possible experimentally available situations involving two-, three- and four-level schemes 
of quadrupole sublevels of the ground and isomeric nuclear states  coupling to one or two coherent fields.  
Specific configurations which offer clear signatures of the isomer excitation advantageous for the more precise experimental 
determination of the transition energy are identified. Furthermore, it is shown that population trapping into the isomeric state 
can be achieved. This paves the way for further nuclear quantum optics applications with $^{229}$Th such as nuclear coherent control.
\end{abstract}


\maketitle

\section{Introduction}

Out of the entire nuclear chart, $^{229}$Th has the so far only known nuclear transition that can be accessed with available optical lasers. The transition energy is at present estimated to be $7.8\pm 0.5$ eV \cite{Bec_09}, corresponding to the vacuum-ultraviolet (VUV) range.  Except for this unique example, direct laser driving of nuclear transitions has been so far discussed in the context of the commissioning of x-ray light sources as the x-ray free electron laser (XFEL) \cite{Kei_PRL06,Pal_PRC08,Wen_PL11}. In both cases, the interest for nuclear quantum optics applications is fueled by the expected parallel to quantum coherence and interference effects in  multilevel atomic and molecular systems driven by 
coherent electromagnetic fields that display intriguing counterintuitive physics and potential important applications \cite{Scullyb, Ficekb}.
Induced atomic and molecular coherences are related to many optical phenomena such as enhanced nonlinear effects \cite{Tewari_PRL_86}, electromagnetically induced transparency (EIT) \cite{Harris90, Boller91}, coherent population trapping \cite{Arimondo}, stimulated Raman adiabatic passage (STIRAP) \cite{Shore}, lasing  without inversion \cite{LWI1,LWI3}, efficient nonlinear frequency conversions \cite{Jain96}, collective quantum dynamics \cite{Haroche, Mihai, Das_PRL08} or vacuum induced coherence in photo-association \cite{Das_PRA12} to name a few. 
The possibility of similar coherent control in nuclear systems,  also related to specific nuclear incentives as 
isomer depletion or a nuclear $\gamma$-ray laser, have been considered with great interest 
\cite{Bal_63,Mar_63,Kei_PRL06,Pal_PRC08}. All these quantum optical 
effects require strong Rabi coupling of the driving field to the considered transitions \cite{Scullyb, Ficekb}. 
Consequently it is important to achieve direct laser driving of nuclear transitions if similar phenomena in nuclei are to be
observed experimentally. 

However, the direct interaction of laser fields with nuclei is generally difficult to achieve due to
(a) small nucleus-laser interaction matrix elements \cite{Mat_98},
(b) ineffective nuclear polarization, as the populations
of the hyperfine levels are usually nearly equal even at very low
temperatures, and (c) the mostly high nuclear excitation energies in the keV to MeV range 
which are only touching the reach of currently available coherent sources. 
In this regard,  the isomeric transition $(I^{\pi}_{is} = 3/2^{+}) \rightarrow (I^{\pi}_{g} = 5/2^{+})$ 
(see Fig. 1) of $^{229}$Th is the notable exception.
This transition is currently a strong candidate for frequency metrology \cite{Pei03, Kaz12} 
and is in focus for several other potential applications like 
temporal variation of the fine structure constant \cite{Fla06} building a nuclear laser in the optical range \cite{Tka11},
 or providing an exciting platform for nuclear quantum optics and coherent control of VUV photons 
 \cite{Kei_PRL06,Wen_PL11,Olga_PRL99,Wen_PRL12,xrayrev13,WenPRC13}. 

Even with energies in VUV range it is at present difficult to  investigate field-induced coherence effects among the 
hyperfine levels of this $^{229}$Th isomeric transition by direct laser coupling. This is primarily due to an $1$ eV uncertainty in the transition energy and a very narrow radiative transition width of $\sim 0.1$ mHz. 
A correspondingly narrow bandwidth VUV laser source is at present only within limitations available and could not be directly employed prior to attaining better knowledge of the transition frequency. 
A viable alternative to attain optical probing of the isomeric transition is to use coherent light scattering off nuclei in 
the low-excitation limit \cite{Han94,Kag_65,Han_99,Rohb,Adamb} by a broadband source like a synchrotron or available VUV lasers. 
In particular, in order to pinpoint the isomeric transition frequency,  a nuclear forward scattering (NFS) setup that takes advantage 
of the coherent light propagation through a solid state sample containing $^{229}$Th nuclei and of the enhanced transient 
fluorescence in the forward direction has been recently proposed \cite{Das_PRL12}. Coherence and interference effects  are expected to help both in the precise determination of the isomeric transition and in developing a platform for nuclear quantum optics studies based on the isomeric transition of $^{229}$Th.

In Ref. \cite{Das_PRL12} a first step in this direction was made by investigating a novel scheme  for the direct measurement of the transition energy via electromagnetically modified NFS involving two fields that couple to three nuclear states. This scheme provides an unmistakable signature of the isomeric  fluorescence in nuclear spectroscopy using  $^{229}$Th nuclei doped in  VUV-transparent crystals. Here we extend the study of coherence and interference effects occurring in the interaction between coherent light and an ensemble of $^{229}$Th nuclei towards two important directions. First, 
we consider a more realistic scenario than in Ref. \cite{Das_PRL12} and take into account the effect of sample inhomogeneities  that broaden the nuclear transition width. This broadening is incoherent and does not enhance the strength oh the light-nucleus coupling; however its effects need to be well understood and controlled for frequency standard applications. So far theoretical studies of the sources of width broadening in an ensemble of $^{229}$Th nuclei consider  the doping $^{229}$Th ion occupying the same location in the crystal unit cell. The theoretical studies in Ref.~\cite{Kaz12} show that in a thorium-doped CaF$_{2}$ crystal, for instance, the thorium ion will most probably replace a Ca$^{2+}$ ion in a Th$^{4+}$ state. The broadenings experienced by the thorium nuclei due to electric and magnetic fields from the surrounding lattice are then assumed to be the same throughout the sample \cite{Rel10,Kaz12,Das_PRL12}.
 This is however difficult to achieve in practice not least due to impurities and color centers, which may occur even during the experiment as irradiation effects. 
Motivated by this, we investigate the effect of different $^{229}$Th doping sites for the NFS response of the irradiated sample. This effect is shown to come in interplay with interference effects due to the hyperfine splitting of the ground and isomeric nuclear states leading to specific beating patterns in the scattered spectra. The effects of different nuclear sites for the fitting of iron $^{57}$Fe M\"ossbauer spectra has been investigated in detail \cite{motif}.

\begin{figure}[!h]
\vspace{-0.4cm}
\scalebox{0.35}{\includegraphics{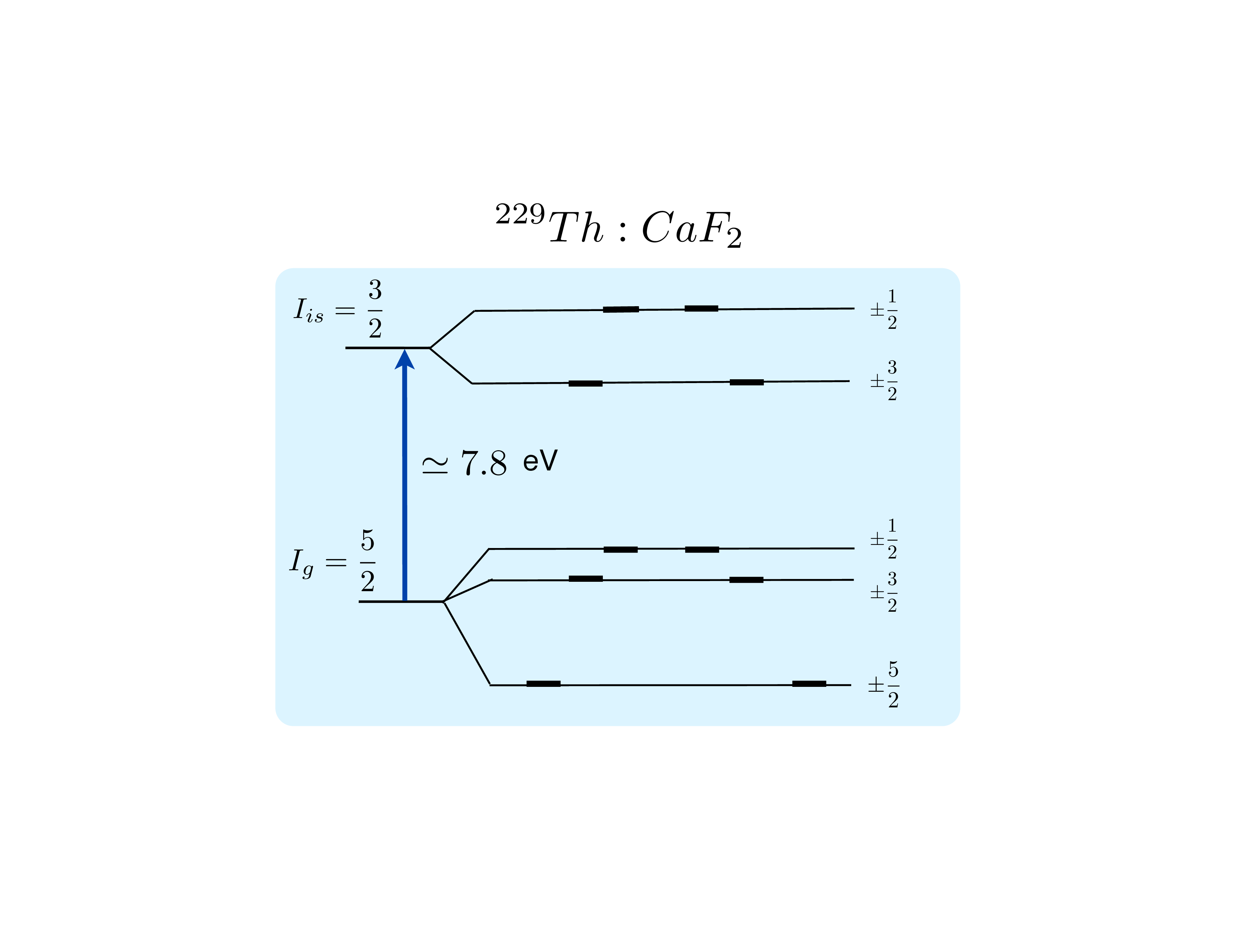}}
  \caption{\label{fig1} Quadrupole  splitting of $^{229}$Th doped in VUV-transparent crystals like 
  LiCaAlF$_{6}$ or CaF$_{2}$. The splittings are on the order of  MHz  \cite{Kaz12}.}
\end{figure}

A second goal of this work is to extend the study of interference effects for the more realistic case of a multilevel quantum system to mimic the quadrupole hyperfine structure of $^{229}$Th presented in Fig.~\ref{fig1}. We investigate a number of possible two-, three- and four-level schemes involving quadrupole sublevels of the ground and isomeric nuclear states  coupling to one or two coherent fields. For the case of three level schemes we consider both the case of two excited hyperfine states coupled to one ground state---the so-called $V$ configuration---and the case of two ground hyperfine levels coupled to one excited state, known as $\Lambda$ configuration, which have been extensively studied in multi-level quantum optical studies in atoms and molecules. 
As driving fields we consider both pulsed and continuous wave coherent VUV sources.  
We find specific configurations  which offer clear signatures of the isomer excitation advantageous for the more precise experimental determination of the transition energy. Furthermore, it is shown that even population trapping into the isomeric state can be achieved in a two-field driven three-level $V$ configuration. This paves the way for further nuclear quantum optics applications with $^{229}$Th.

The paper is organized as follows. In section II we describe the model and discuss the basic properties of coherent forward 
scattering of VUV laser from an ideal system---an ensemble of two level nuclei doped in a VUV-transparent crystal lattice 
environment. We then develop a general theoretical framework in terms of the Maxwell-Bloch formalism to study coherent forward 
scattering of VUV from realistic systems which includes decoherence, multiple resonances due to multilevel structure of nuclei and 
inhomogeneity due to different doping sites. In section III we use this general theoretical framework to study the effect of 
environment-induced decoherence (for instance via spin-lattice relaxation) and different doping sites on the NFS signal 
for the otherwise analytically solvable, simple model of two level nuclei doped in a VUV-transparent crystal lattice environment. 
In section IV we consider an ensemble of multi-level nuclei doped in a crystal lattice environment
in several different configurations driven by either one or two optical fields. In section IV A we study NFS of a
VUV pulse from a three level scheme in which one hyperfine level of the isomeric state is coupled to two different hyperfine 
levels of the nuclear ground state by a probe and a control field. We consider both continuous wave and pulsed fields as control field  and discuss the corresponding quantum interference signatures in the NFS spectra. The effect of 
different $^{229}$Th doping sites on the NFS signal in such a three level setup is investigated. In section IV B we study 
NFS of a VUV pulse from a three level scheme in which two hyperfine levels of the isomeric state are coupled to a common hyperfine level of the nuclear ground state by a single optical field. We also investigate in this model the effect of coherence created between the hyperfine levels of the isomeric state on the population dynamics. 
In section IV C we then study NFS of a VUV pulse driving a four level configuration with two near degenerate 
transitions.  Finally  we conclude  in section V with a summary of our findings.

\section{Model and Theoretical Framework}
We begin our investigation by studying NFS off an ensemble $^{229}$Th nuclei doped in a 
crystal lattice environment. Nuclear spectroscopy using  $^{229}$Th nuclei doped in  VUV-transparent crystals such as  LiCaAlF$_{6}$ \cite{Rel10} or CaF$_{2}$ \cite{Kaz12} offers the possibility to increase the  nuclear excitation  probability significantly due to the high doping density of up to $10^{18}$ Th/cm$^3$. Both  LiCaAlF$_{6}$ and CaF$_{2}$ have large band gaps and present good transparency at the probable transition wavelength, such that the interplay with 
electronic shells in processes such as the electronic bridge \cite{eb10}, internal conversion \cite{newprl} or nuclear excitation via electron capture or transition \cite{NEECT} can be neglected. When confined to the Lamb-M\"ossbauer regime of recoilless nuclear transitions, the excitation of the isomeric transition will occur coherently in the forward direction, leading to a speed-up of the initial nuclear decay depending primarily on the sample optical
thickness. These effects are well known from NFS of synchrotron radiation \cite{Smi_96,Bur_99} driving
M\"ossbauer nuclear transitions in the x-ray regime, and have been addressed for the first time for the case of $^{229}$Th in Ref. \cite{Das_PRL12}. 

As nuclear level scheme  we use the quadrupole structure of $^{229}$Th with hyperfine level energies given by 
$E_{m} \simeq Q_{is(g)}(1-\gamma_{\infty})\phi_{zz}[3m^{2}-I_{is(g)}(I_{is(g)}+1)]/
[4I_{is(g)}(2I_{is(g)}-1)]$, where $Q_{is(g)}$ = 1.8 eb (3.15 eb) is the quadrupole 
moment of the isomeric (ground) level, $\gamma_{\infty} = -(100-200) $ is the antishielding factor 
and $(1-\gamma_{\infty})\phi_{zz} \sim -10^{18}$ V/cm$^{2}$ electric field gradient \cite{Tka11,Kaz12}. 
Fig.~\ref{fig1} shows the energy scheme of $^{229}$Th with the electric quadrupole splitting \cite{Kaz12} 
of the ground and excited $^{229}$Th nuclear states of spins $I_g$=5/2 and $I_{is}$=3/2, respectively. We 
use the recently proposed $^{229}$Th:CaF$_{2}$ crystal \cite{Kaz12} parameters for numerical
evaluation of the NFS signal. 

A radiation pulse denoted in the 
following as probe driving the relevant nuclear transition shines perpendicular to the nuclear sample and the scattering response in the forward direction is recorded. Depending on the pulse polarization, different hyperfine transitions will be driven. 
This setup follows the typical NFS experiments  extensively performed with $^{57}$Fe \cite{Kag_99, Han_99}.

\section*{Analytical framework: two level approximation}
We consider first the simplest possible model for NFS study---two level nuclei interacting with an incident VUV 
laser pulse. In case of $^{229}$Th, such a two level system can be formed by selectively driving a $m_{e}-m_{g} = 0, \pm 1$ 
magnetic dipole transition with a VUV pulse using a cooled sample where not all hyperfine ground states are populated. Here $m_{e}$ and $m_{g}$ denote the projections of the excited and ground state nuclear spins on the quantization axis, respectively.  The VUV laser pulse may be generated via nonlinear sum-frequency mixing 
\cite{Irrg98}, or a harmonic of a VUV frequency comb \cite{Jones05, Ozawa08} around the isomeric wavelength.

In NFS  the resonant scattering off the nuclear ensemble 
occurs via an excitonic state---an excitation coherently spread out over a large number of nuclei. 
When the scattering is coherent, the nuclei return to their initial state  erasing in the process 
any information of the scattering path. This leads to cooperative emission with scattering only in 
the forward direction (except for the case of Bragg scattering \cite{Han_99,Smi_96,Kag_99}) and decay rates modified by the formation 
of sub- and superradiant modes of emission. Coupled with the narrow line-width of nuclear transition, this co-operative 
feature of NFS has been exploited in studying co-operative Lamb shifts \cite{Scully2007,Roh_Sci10}, single-photon entanglement generation in 
x-ray regime \cite{Ada_PRL09, Ada_JMO10}, storage and modulation of single hard x-ray photons \cite{Wen_PRL12} and coherent optical scheme of direct 
determination of the $^{229}$Th isomeric transition \cite{Das_PRL12}.
The time evolution of the forward scattering response exhibits pronounced intensity modulations characteristic of the coherent 
resonant pulse propagation \cite{Bur_99,Cri_PRA70,Kag_79}. This modulation is known in nuclear condensed-matter physics
under the name of dynamical beat and, for a single resonance (two level approximation)
and a short $\delta(t)$-like exciting pulse, has the form
\begin{equation}
\label{eq0}
E(t) \propto \xi e^{-\tau/2}J_{1}(\sqrt{4\xi\tau})/\sqrt{\xi\tau} 
\end{equation}
where $E$ is the transmitted pulse envelope, $\tau$ a dimensionless time parameter
$\tau = t/t_0$ with $t_0$ denoting the natural lifetime of the nuclear
excited state, $\xi$ is the optical thickness and $J_1$ is the Bessel
function of the first kind. The optical thickness is defined
as $\xi = N\sigma L/4$ \cite{Han_99}, where $N$ is the number density of
nuclei, $\sigma$ is the nuclear resonance cross-section, and $L$ is the
sample thickness. From the asymptotic behavior of the Bessel function of
first kind, we see that for early times of evolution $\tau << 1/(1+\xi)$ immediately
after the excitation pulse ($\tau = t = 0$), the
response field has the form $E(t) \propto \xi \exp[-(1+\xi)\tau/2]$
showing the speed-up of the initial decay by $\xi$. At later times, the decay becomes subradiant, i.e., with a
slower rate comparable to the incoherent natural decay rate due to destructive interference between radiation
emitted by nuclei located at different depths in the sample. 

In the case of $^{229}$Th doped in VUV transparent crystals,
both the conditions for recoilless, coherent excitation and
decay and broadband excitation are fulfilled for scattering
in the forward direction. The incident VUV laser pulse 
duration is much shorter than the nuclear lifetime and 
provides broadband excitation. Since the crystal is 
expected to be transparent at the nuclear transition
frequency \cite{Kaz12}, the main limiting factor for coherent
pulse propagation, namely, electronic photoabsorption, is
not present in this case. Thus we can expect pronounced intensity modulations 
characteristic of the coherent resonant pulse propagation given by
the analytical expression (\ref{eq0}) for the forward scattered 
field from an ensemble of two level $^{229}$Th nuclei. 
\section*{General framework: Maxwell-Bloch formalism}
Instead of using Eq. (\ref{eq0}) to study the behavior of the scattered field intensity 
in the forward direction,  we take an alternative  approach 
and numerically evaluate the Maxwell-Bloch equations \cite{Scullyb}. 
This allows to include multiple transitions between hyperfine splitting levels in $^{229}$Th (this will become important 
in later sections when we consider multiple levels) and 
consider decoherence processes like inhomogeneous broadening occurring due to spin-spin relaxation. 
The interaction Hamiltonian for the system of multi-level nuclei in the dipole approximation and in a frame rotating with 
the frequency of the incident laser is given by
\begin{equation} 
\label{eq1}
\mathcal{H}  =  \hbar\sum_{j}\Delta_{j} S^{-}_{j}S^{+}_{j}-\sum_{j}\left(\frac{\Omega_{j}}{2}(\vec{\varepsilon}_{fj}\cdot\vec{\varepsilon}_{nj})S^{+}_{j}+H.c.\right)
\end{equation}
where $\Delta = \omega_{j}-\omega_{L}$ is the so-called detuning with $\omega_{j}, \omega_{L}$ being the $j$th nuclear transition and the driving laser 
frequency, respectively. Here $\Omega_{j}$ is the space- and  time-dependent Rabi frequency of the driving field for the nuclear transition of interest labeled by $j$ and 
$\vec{\varepsilon}_{fj}, \vec{\varepsilon}_{nj}$ are the polarizations 
of the incident light and of the transition, respectively. Furthermore, $S^{+}_{j}(S^{-}_{j})$ are the nuclear raising (lowering) 
operators for the $j$th transition in analogy to the atomic raising (lowering) operators and satisfying the $SU(2)$ 
angular momentum algebra. 

The interaction of the doped nuclei with their environment consisting of atoms, electrons and 
nuclei of other species in the VUV crystals leads to relaxation and decoherence in the system. 
The major mechanisms of relaxation and decoherence are the spontaneous decay---in the crystal assumed to occur 
only radiatively---and the spin relaxation of thorium nuclei due to interaction with the random magnetic field created 
by the surrounding fluorine spins in CaF$_{2}$ \cite{Kaz12}.
Mathematically the effect of population relaxation can be included in the formalism by the following Louvillian operators
\begin{equation}
\label{eq2}
\mathcal{L}\rho^{(\alpha)}  = -\sum_{j}\frac{\gamma_j}{2}(S^{+}_{j}S^{-}_{j}\rho^{(\alpha)}+\rho^{(\alpha)} S^{+}_{j}S^{-}_{j}-2S^{-}_{j}\rho^{(\alpha)} S^{+}_{j})\\
\end{equation}
where $\gamma_j$ is the population relaxation rate of the $j$th transition and $\rho^{(\alpha)}$ 
is the density operator for the nuclei of type $\alpha$. The relaxation rate for a relevant 
nuclear hyperfine transition  can be related to the total decay rate with the help of  
the corresponding  Clebsch-Gordon coefficient  \cite{Pal_PRC08}. 
In practice, due to more than one  doping sites
of the VUV crystal, the nuclear ensemble  might consist of $^{229}$Th nuclei with different hyperfine splittings, which can in 
principle be treated as different species of nuclei doped throughout the sample. 
The index $\alpha$  will be used to differentiate between the nuclei doped in different nuclear sites. 

In order to obtain the NFS time spectra, we evaluate the net generated intensity 
$I(z,t) = |\Omega(z,t)|^2$ at the exit from the medium. The behavior of the output 
field $\Omega(z,t)$ is given by the Maxwell equations involving the induced nuclear 
currents which in turn are related to the coherence terms.  The coherence in 
the system can be found by studying the dynamics of the density matrix $\rho^{(\alpha)}$. 
This thus leads to the coupled Maxwell-Bloch equations \cite{Scullyb}:
\begin{equation}
\label{eq3a}
\partial_{t}\rho^{(\alpha)} =  \frac{1}{i\hbar}\left[ \mathcal{H},\rho^{(\alpha)}\right]+\mathcal{L}\rho^{(\alpha)}+\mathcal{L}_{d}\rho^{(\alpha)},
\end{equation}
\begin{equation}
\label{eq3b}
\partial_{z}\Omega_{j}  +  \frac{1}{c}\partial_{t}\Omega_{j} =  i \sum_{\alpha}\sum_{lk}\eta^{(\alpha)}_{lk}a_{lk}\rho^{(\alpha)}_{lk},
\end{equation}
where $a_{lk}$ is the Clebsh-Gordon coefficient, $\eta^{\alpha}_{lk} = \Gamma_{0}\xi^{(\alpha)}_{lk}/2L$ for the transition between the states $l \rightarrow k$, $(l\neq k)$ and $\Gamma_{0}$ 
is the natural (radiative) decay rate for the nuclei.  Note that the optical thickness $\xi$ varies with the group index $\alpha$ and 
the transitions due to its dependence on the doping density and transition line-width via the nuclear resonance cross-section \cite{Andr12}. 
For typical $^{229}$Th-doped VUV crystal parameters $\Gamma_{0}\sim 0.07$ mHz, $\eta = 100$ Hz/cm, where $\xi = 10^{6}$ and $L = 1$ cm \cite{Bec_09, Kaz12, Das_PRL12}. 
The term $\mathcal{L}_{d}\rho^{(\alpha)}$ in the above equations represents the decoherence of the relevant 
nuclear transitions for both few and multi-level nuclei. 
\section{Coherence effects in NFS off an ensemble of two level nuclei}
%
\begin{figure}[!b]
\vspace{-0.4cm}
\scalebox{0.25}{\includegraphics{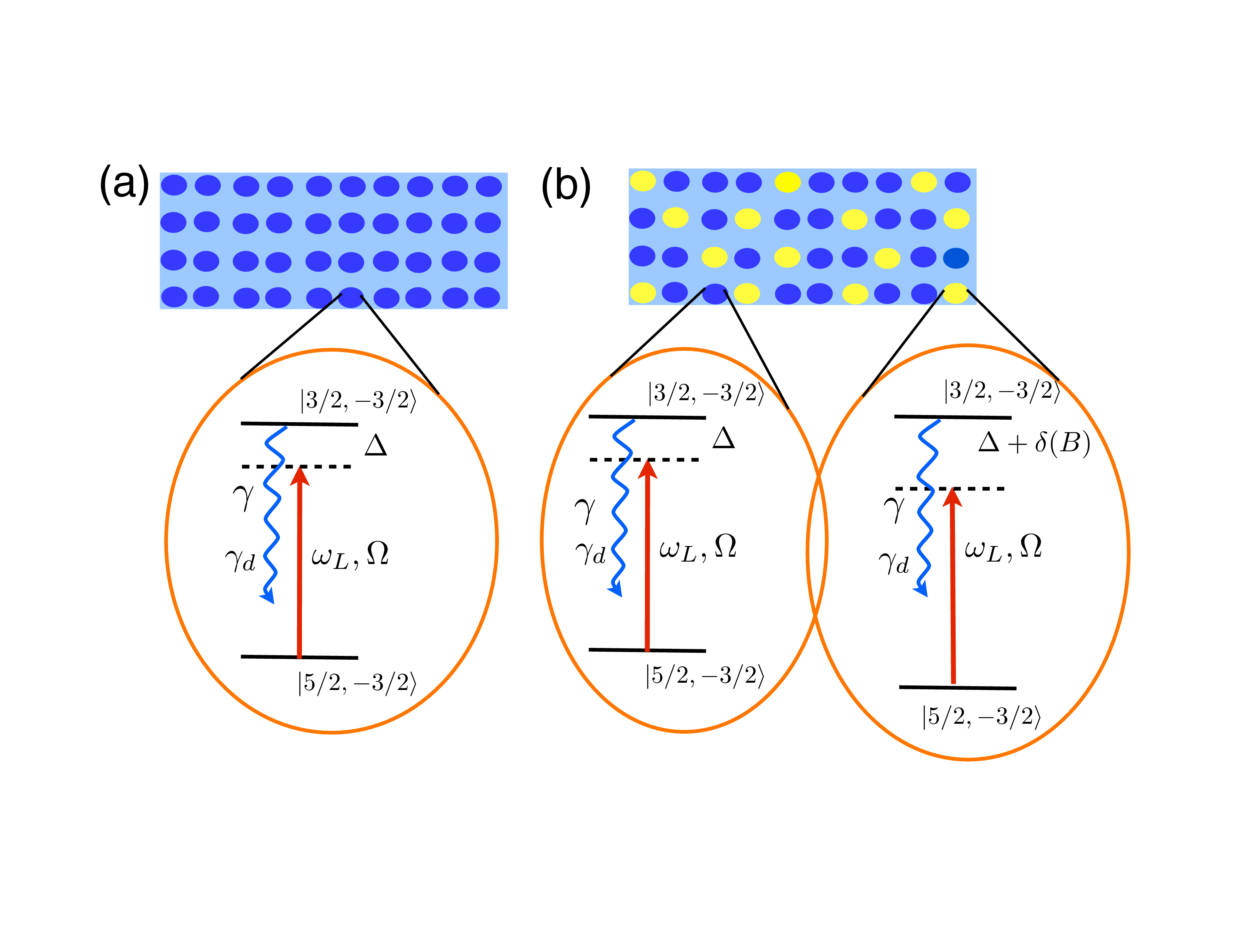}}
  \caption{\label{fig2} (Color online) (a) Level scheme for VUV excitation of the doped $^{229}$Th in the crystal lattice. 
  All doping thorium nuclei share the same environment and lattice-generated fields. 
  (b) The same for two different doping sites. The darker (blue) thorium nuclei are influenced by a different environment 
  than the lighter (yellow) ones. See text for further explanations.}
\end{figure}
In this section we investigate the properties of NFS of a VUV laser pulse off an ensemble of simple two level $^{229}$Th doped in a VUV-crystal 
lattice environment. This problem was discussed earlier in section II  in an analytical framework. Here, we use the general framework 
developed in the previous section to numerically simulate the additional effect of decoherence and different doping sites on the NFS for Th-doped VUV-transparent crystals. 

We begin our investigation by assuming the exciting VUV pulse to be linearly polarized
and for now driving resonantly a single transition. 
Thus the hyperfine manifold of the $^{229}$Th nuclei can be approximated by a two level nuclei satisfying the selection rule $m_{e}-m_{g} = 0$. 
For our model calculation we explicitly choose the levels to be $\{|3/2,-3/2\rangle = |1\rangle , |5/2,- 3/2\rangle = |2\rangle\}$ as shown in 
Fig \ref{fig2}(a). Thus in the two level situation the number of transitions involved is only one and the 
nuclear raising lowering operators are defined as $S^{+} = |1\rangle\langle 2|,S^{-} = (S^{+})^{\dagger}$. The corresponding 
field equation (\ref{eq3b}) is thus driven by a single coherence term $\rho^{(\alpha)}_{12}$ with the
subscript $j$ on $\Omega$ replaced by $p$ denoting the probe field. 
Furthermore, the decoherence term $\mathcal{L}_{d}\rho^{(\alpha)}$ in Eq.~(\ref{eq3a}) for the two level configuration can be written as
\begin{equation}
\label{eq3c}
\mathcal{L}_{d}\rho^{(\alpha)}  =  -\frac{\gamma_{d}}{2}(S^{z}S^{z}\rho^{(\alpha)}+\rho^{(\alpha)} S^{z}S^{z}-2S^{z}\rho^{(\alpha)} S^{z}),
\end{equation}
where $\gamma_{d}$ is the decoherence rate of the transition and 
$S^{z} = 1/2(|1\rangle\langle 1|-|2\rangle\langle 2|)$ is the energy operator.
For the present analysis we assume that all thorium nuclei are occupying the same doping site in the VUV crystal, i.e., all dopants  
experience the same lattice fields and hence we can drop the
index $\alpha$. The coupled Maxwell-Bloch equations (\ref{eq3a}) and (\ref{eq3b}) are solved taking into account  the 
initial condition that all the population is in the ground state ($\rho_{22}(0) = 1$). 
The exciting VUV pulse is assumed to have a Gaussian shape defined 
by the initial and boundary conditions,
\begin{equation}
\label{eq4}
\Omega_{p}[0,t] = \Omega_{p0}\exp\left[-\left(\frac{t-t_{0}}{\tau}\right)^2\right] , \ \ \Omega_{p}[z,0] = 0.
\end{equation} 
The NFS time spectrum after the passage of the excitation pulse is shown by the dotted line in Fig.~\ref{fig3}
for a resonant  probe field and with a decoherence (spin relaxation) rate of $\gamma_{d} = 2\pi\times108$ Hz \cite{Kaz12}. A Rabi frequency of 
$\Omega_{0} = 10^{6}\Gamma_{0}$ is assumed for the numerical computation of the Maxwell-Bloch 
equations in the present case. Our result enforces the earlier found behavior of NFS spectra of $^{229}$Th in such a setup \cite{Das_PRL12}.
The nuclear forward scattering time spectra are not sensitive to the laser detuning and have a behavior similar to 
that in Fig. ~\ref{fig3} for a large range of detuning $\Delta$ ($0 - 1$ KHz). Furthermore, the slope of the time spectra 
essentially follows the decoherence rate $e^{-2\gamma_{d}t}$ of the two level nuclear system. 

\subsection*{Different  doping sites}
In the above case we have assumed that any energy shifts or broadening induced by the crystal lattice
(like hyperfine interactions) are the same for all thorium nuclei. However, in practice the perfectly doped crystal with 
all thorium nuclei in the same doping site is hard to 
achieve, since impurities and color centers are often also present.  The energy splittings of 
the driven hyperfine transitions  can therefore be different at different lattice sites. 

We mimic this situation for a simple case by considering the VUV crystal with  two groups of dopant nuclei with different 
splitting of the isomeric transition energy. In the Hamiltonian (\ref{eq1}) the index $j$ takes the values $j = 1,2$ now, 
with $\Delta_{1} = \Delta$ and  $\Delta_{2} = \Delta_{1}+\delta(B)$
representing the second group of nuclei depicted by yellow circles in Fig. \ref{fig2}(b). Here $\delta(B)$ is an additional 
intrinsic magnetic field-dependent detuning due to the different hyperfine splitting of the second group of nuclei. 
The incident field in this case interacts with both groups of  dopant nuclei leading to simultaneous contribution 
of the nuclear coherences from each group in the generated signal. The field equation in the Maxwell-Bloch formalism 
change accordingly to incorporate the contribution of coherences from both nuclear sites. In Eq. (\ref{eq3b})
the index $\alpha = 1, 2$ stands now for the two groups of doping thorium nuclei. The dynamics of the system follows
from Eq. (\ref{eq3a}) with the modified Hamiltonian as discussed above and the corresponding population relaxation and 
decoherence given by Eqs. (\ref{eq2}) and  (\ref{eq3c}). For simplicity we have considered equal doping density of the 
groups such that $\eta^{(1)} = \eta^{(2)}$.
\begin{figure}[!t]
\vspace{0.5cm}
\scalebox{0.8}{\includegraphics{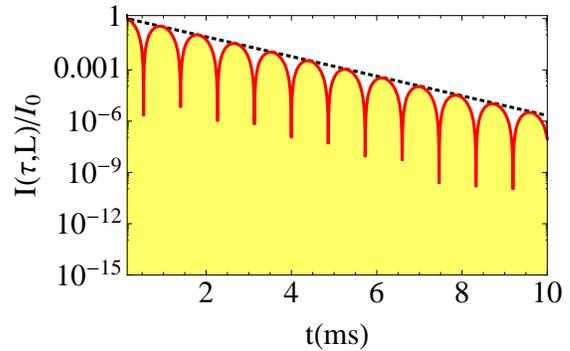}}
  \caption{\label{fig3} (Color online) NFS time spectra scattered off an ensemble of two level nuclei. 
  The dotted (black) line shows the results for single site doping 
  while the solid (red) one with yellow shading is for two site doping, respectively. Here $\gamma$=0.356$\Gamma_{0}$, 
  $\eta^{(1)} = \eta^{(2)} = \eta/2 \sim 91$ Hz/cm for $\xi = 10^{6}$ 
  with the pulse and detuning parameters $\Omega_{0} = 10^{6}\Gamma_{0}$, $t_{0} = 0.1$ ms, $\tau = 0.01$ ms, and 
  $\delta(B) = 10^{(8)}\Gamma_{0}$, respectively.}
\end{figure}

The NFS time spectrum for two different doping sites is shown in Fig. \ref{fig3} by the solid line with yellow shading. We have assumed 
resonant driving of the transition for the first group of nuclei $(\Delta = 0)$ and $\delta(B) = 10^{8}\Gamma_{0}$ 
corresponding to an intrinsic magnetic field of $\sim 100$ Gauss. The behavior of the time spectra is distinctly modified in the presence of 
such non-uniform doping. We find quantum beats in the NFS spectra of the order of $\delta(B)$
owing to inter-nuclei quantum interference between the two groups with different hyperfine splittings. 
The beating pattern is influenced by the decoherence in the system and has an envelope governed
by the decay rate $e^{-2\gamma_{d}t}$. 

Unfortunately, in experiments such spectra cannot be differentiated from background
coming from other unwanted electronic processes in the VUV crystal that can be active at non-resonant probe laser 
frequencies. The NFS signal lacks any distinctive signature of nuclear resonance, particularly for 
uniform doping of the sample, being insensitive to even large detunings.  
In the following sections we thus move on to some more complicated and realistic multi-level models 
of $^{229}$Th in a search for isomeric signatures provided by coherence and quantum interference features in the scattered light. 

\section{ Coherence effects in NFS off an ensemble of multi-level nuclei} 
In this section we systematically study the effect of quantum coherence and interference on the NFS signal emitted from 
an ensemble of multi-level nuclei. In the first part of this section we consider a nuclear ensemble interacting with two fields which selectively drive two magnetic dipole transitions. The level configuration is such that in each nucleus there is one upper level connected 
to two lower levels by the two fields forming a $\Lambda$ system. The $\Lambda$ scheme is a typical set for interference effects  and has been quite extensively studied in atomic quantum optics during the past decade. In the second part of the section we consider the inverse level configuration of the nuclear ensemble, i.e., a $V$ configuration 
where a single field drives two transitions formed by two upper and one lower nuclear levels. A second additional field that 
couples the upper states is also considered in this case to create coherence among the latter. Finally in the last part 
 we study a nuclear ensemble in which a single field drives two nearly degenerate transitions formed by four levels in a two-upper and two-lower states configuration.

\subsection{Nuclear ensemble in a three level $\Lambda$-configuration}
\begin{figure}[!h]
\scalebox{0.24}{\includegraphics{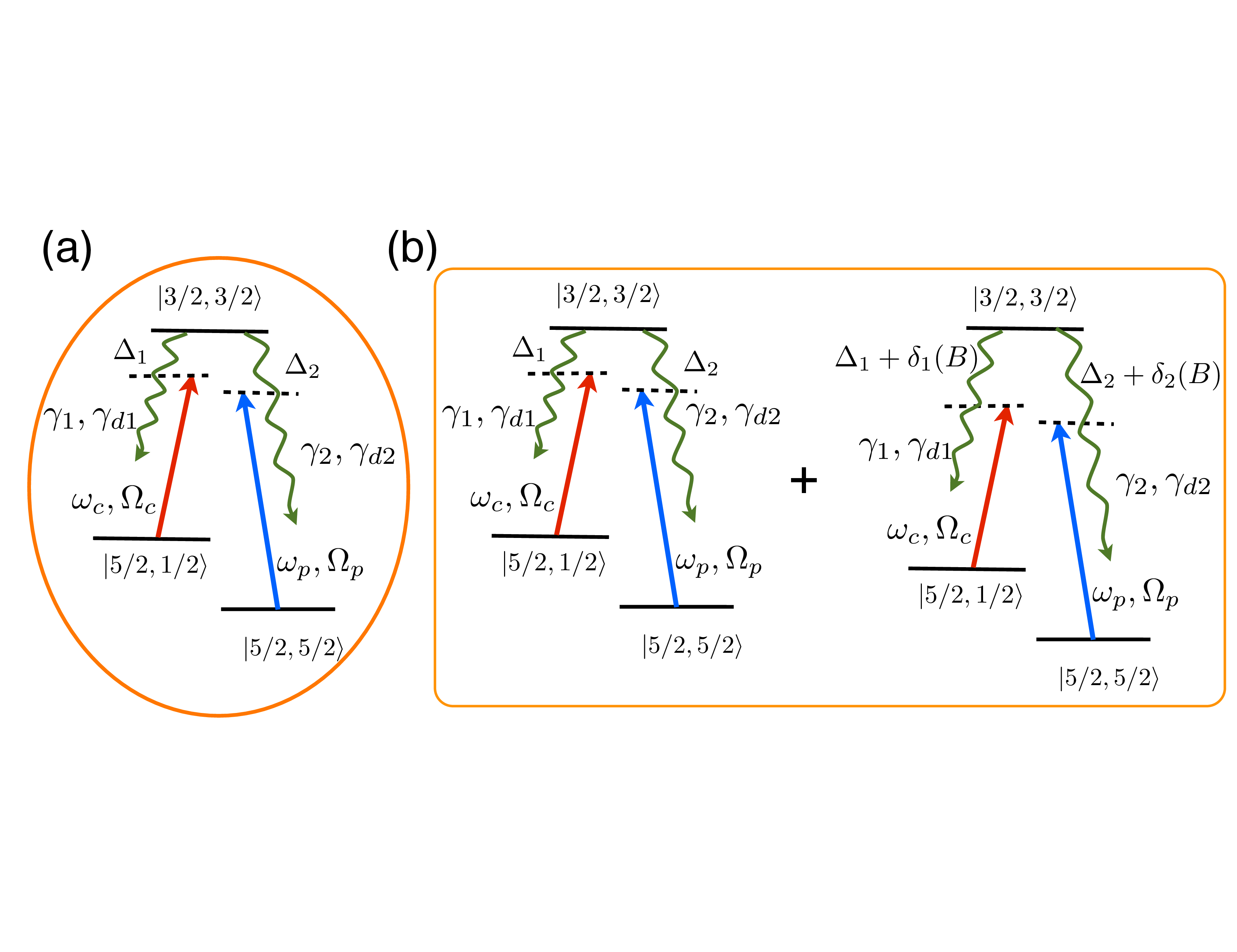}}
  \caption{\label{fig6} (Color online) Level schemes of the doped $^{229}$Th in the crystal lattice exposed to  two VUV fields. (a) The doping 
  occurs in one site only and thus all the thorium nuclei experience the same environment.  (b) Two-site doping 
with different hyperfine splittings due to different magnetic environments in the crystal lattice. See text for further explanations.}
\end{figure}

In this section we consider an ensemble of nuclei occupying the same doping site in a VUV crystal interacting with two 
electromagnetic fields. One of the fields is a left-circularly polarized weak 
VUV pulse denoted in the following as probe driving the $m_{e}-m_{g} =-1$ magnetic dipole transition, while the other is a strong right circularly polarized continuous wave 
(cw) denoted as control driving the $m_{e}-m_{g} = 1$ transition. 
Due to the selected polarizations, the two fields couple two nuclear ground states to a common excited state 
forming a $\Lambda$-type scheme.
Such a $\Lambda$-configuration of the nuclear levels can be achieved in $^{229}$Th by driving the transition 
$|5/2,\pm 1/2\rangle \rightarrow |3/2,\pm 3/2\rangle$ with the control laser and the 
$|5/2,\pm 5/2\rangle \rightarrow |3/2,\pm 3/2\rangle$ transition with the probe pulse as shown in Fig. \ref{fig6}(a). 
Note that recently such a three level two field scheme was proposed for the coherence-enhanced optical determination the 
isomeric transition energy to a high precision \cite{Das_PRL12}. 

For the three-level scheme, the interaction Hamiltonian is given by Eq. (\ref{eq1}) with  $j = 1, 2$ representing 
the transitions $|5/2,\pm 1/2\rangle \rightarrow |3/2,\pm 3/2\rangle$ and 
$|5/2,\pm 5/2\rangle \rightarrow |3/2,\pm 3/2\rangle$, respectively. The nuclear raising (lowering) 
operators are defined in a three-level basis $\{|3/2,\pm 3/2\rangle = |1\rangle, |5/2,\pm 1/2\rangle = |2\rangle,|5/2,\pm5/2\rangle = |3\rangle\}$
as $S^{+}_{1} = |1\rangle\langle 2|$, $S^{+}_{2} = |1\rangle\langle 3|$, $S^{-}_{k} = (S^{+}_{k} )^{\dagger}$, $k = 1,2$.
In the following we consider the transitions between the states with positive angular momentum projection $|5/2, 1/2\rangle$, $|5/2, 5/2\rangle$ and $|3/2, 3/2\rangle$; similar results apply when the fields couple the states with negative ones.
There are two Rabi frequencies $\Omega_1 = \Omega_c$ and $\Omega_2 = \Omega_p$ in the Hamiltonian 
corresponding to the control and probe fields with the detunings $\Delta_{1} = \omega_{12}-\omega_{c}$ and $\Delta_{2} = \omega_{13}-\omega_{p}$, 
where $\omega_c$, $\omega_p$ and $\omega_{12}$, $\omega_{13}$ are the frequencies of the control, probe and the relevant nuclear transitions, respectively. 
The population relaxations in the three level system are given by the Louvillian operator in Eq. (\ref{eq2}) with $ j = 1, 2$ 
and $\gamma_{1}$ and $\gamma_{2}$ being the  relaxation rates of the isomeric state to the two 
ground levels $|5/2, 1/2\rangle$ and $|5/2,5/2\rangle$, respectively. All  nuclei 
 experience the same environment and   hyperfine splitting  such that we drop the index $\alpha$ on the density operator.
The decoherence of the transitions are incorporated in the dynamical equations by means of the decoherence matrix,
\begin{equation}
\label{eq7}
\mathcal{L}_{d}\rho= -\left[ 
\begin{array}{cccc}
  0 & \gamma_{d1}\rho_{12} & \gamma_{d2}\rho_{13}\\
  \gamma_{d1}\rho_{21} & 0 & 0\\
  \gamma_{d2}\rho_{31} & 0 & 0
\end{array}  
\right]
\end{equation}
where $\gamma_{d1}$ and $\gamma_{d2}$ are the decoherence rates of the transitions 
$|5/2, 5/2\rangle \rightarrow |3/2, 3/2\rangle$ and $|5/2, 1/2\rangle \rightarrow |3/2, 3/2\rangle$, respectively.
We consider the system to be closed, thus enforcing no population relaxation of the ground states (no leakage of population 
from ground to other hyperfine states) in our model. 

The Maxwell-Bloch equations for the three level scheme are given by Eqs. (\ref{eq3a}) and (\ref{eq3b}) 
with $ j = p,\  l = 1, \ k = 3$ in Eq.  (\ref{eq3b}) and without the summation over index $\alpha$.
The coherence $\rho_{13}$ that is the source term in 
the field equation is evaluated from the Bloch Eq. (\ref{eq3a}). 
The second term of the Maxwell-Bloch equations only involves the probe field as the control 
field is assumed to be continuous wave. Time spectra of the NFS are numerically evaluated as the 
scattered intensity of the probe field from the Maxwell-Bloch equations. 

\begin{figure}[!h]
\scalebox{0.8}{\includegraphics{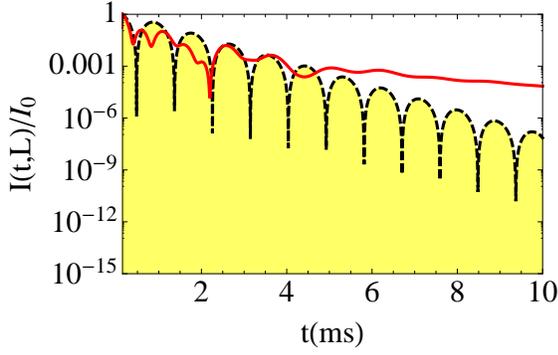}}
  \caption{\label{fig7} (Color online) The time domain spectrum of NFS for a $\Lambda$ three level configuration with two fields setup. 
  The black dotted (solid red) curve with yellow shaded region represents the scattering signal for single (two site) doping. 
 Here we have considered equal doping density for the two nuclear sites such that $\eta^{(1)}_{13} = \eta^{(2)}_{13} = \eta_{13}/2 \simeq  227$~Hz/cm for 
 $\xi \simeq 10^6$. The additional detuning introduced by the two different doping sites is  
  $\delta_{i}(B) = 10^{8}\Gamma_0$. The spontaneous decay rates of the transitions are respectively, 
  $\gamma_{1} = 0.088\Gamma_{0}$ and $\gamma_{2} = 0.889\Gamma_{0}$. The control and probe fields 
  are equally detuned from the level $|3/2, 3/2\rangle$ by $\Delta = 10^{5}\Gamma_0$. The control 
  field is assumed to have a Rabi frequency of $10^{8}\Gamma_0$.}
\end{figure}

For numerical computation we have considered the initial condition $\rho_{33}(0) = 1$ for the 
Bloch equations. The initial and boundary conditions on the probe field are given by Eq. (\ref{eq4}) 
with $\Omega_{p0} = 10^{6}\Gamma_{0}$. The Rabi frequency of the control field is assumed to be
$10^{2}\Omega_{p0}$ which is of the order of MHz. Furthermore, as the quadrupole splitting 
of the thorium ground state can be determined experimentally \cite{Kaz12}, following \cite{Das_PRL12} 
we set the detunings of the control and probe fields to be identical, i.e., $\Delta_1= \Delta_2 = \Delta$. 
The time spectrum of scattered intensity in forward direction are shown in Fig. (\ref{fig7}) by the dashed black curve 
with yellow shading  for decoherence rates $\gamma_{d1} =  2\pi\times158$ Hz, $\gamma_{d2} =  2\pi\times 251$ Hz 
and detuning $\Delta = 10^{5}\Gamma_{0}$. The spectrum as seen from the figure is quite different 
to that of the two level case with uniform doping. Here we observe definite signatures of nuclear excitation in 
the form of quantum beats in the NFS signal \cite{Han_99,Das_PRL12}. Furthermore, the beats are influenced by decoherence 
on the probe transition and have an envelope governed by the rate $e^{-2\gamma_{d2}t}$.

The existence of the quantum beats in such a three level configuration can be explained in the 
dressed-state  picture \cite{CohenB}. The strong control field dresses the transition 
$|5/2,1/2\rangle \rightarrow |3/2,3/2\rangle$ to form two new dressed states 
$|\Psi_{+}\rangle$ and $|\Psi_{-}\rangle$ which are linear combinations of the states 
$|3/2, 3/2\rangle$ and $|5/2, 5/2\rangle$ such that 
\begin{eqnarray}
\label{eq8}
|\Psi_{+}\rangle & = & \sin\theta |5/2, 1/2\rangle +\cos\theta |3/2, 3/2\rangle\nonumber\\
|\Psi_{-}\rangle & = & \cos\theta |5/2, 1/2\rangle -\sin\theta |3/2,3/2\rangle
\end{eqnarray}
where $\tan(2\theta ) = \Omega_{c}/\Delta$. The incoming probe excitation then 
couple these two dressed states to the $|5/2, 1/2\rangle$ level. Thus the nuclear resonance driven 
by the probe field is split into a doublet \cite{Shvydko99} via the Autler-Townes effect \cite{Townes55} 
forming two transition pathways for emission. The quantum interference for emission along
these two transitions creates the beating pattern with the dressed frequency $\tilde{\Omega}$, where
$\tilde{\Omega} = \sqrt{\Delta^{2}+\Omega^{2}_{c}}$. As in the present case $\Delta << \Omega_{c}$, the dressed 
frequency is approximately the same as the Rabi frequency of the control field. Note that the beating pattern found in 
this case is similar to the two level situation with two-site doping. This is due to the fact that in both cases 
the NFS involves two transition pathways with frequency difference of $10^{8}\Gamma_{0}$. The physical origin of the quantum beats is however different. While the beats arise due to 
intra-nuclei quantum interference in three level case with uniform doping, for the two-level case with two doping sites  
they come about as a result of inter-nuclei quantum interference.

So far we have assumed for the three-level system only one doping site. For a two-doping site situation
the Hamiltonian description of the $\Lambda$-configuration now comprises of two 
Hamiltonians of the form (\ref{eq1}) for group $1$ and $2$ of doping nuclei, 
respectively. The detuning $\Delta_j$ in the Hamiltonian for the second group however has now 
additional terms, $\Delta_{1} \rightarrow \Delta_{1} + \delta_{1}(B)$ and 
$\Delta_{2} \rightarrow \Delta_{2} + \delta_{2}(B)$. Here $\delta(B)$ is an additional intrinsic 
magnetic field $B$ dependent detuning due to the different hyperfine splitting of the second group of nuclei.  
The probe VUV pulse will now simultaneously interrogate both doping groups, with the field equation 
having nuclear coherence contributions from  both doping sites  $1$ and $2$. The index $\alpha$ on the density
operator and $\eta$ now takes the values $\alpha = 1, 2$ representing the nuclei in the two  doping sites. 
The field equation for propagation in this case is given by Eq. (\ref{eq3b}) with the summation running over different 
values of $\alpha$, 
\begin{equation}
\label{eq9}
\partial_{z}\Omega_{p} + \frac{1}{c}\partial_{t}\Omega_{p} =  i a_{13}(\eta^{(1)}_{13}\rho^{(1)}_{13}+\eta^{(2)}_{13}\rho^{(2)}_{13}).
\end{equation}
The dynamics of the two groups of nuclei follows from Eq. (\ref{eq3a}) with $\alpha = 1, 2$ and the modified Hamiltonian 
as discussed above. The corresponding population relaxation and decoherence are given by Eqs. (\ref{eq2}) and (\ref{eq7})
for each $\alpha$. We assume for computational purposes $\delta_{1} (B) = \delta_{2}(B)  = \delta = 10^{8}\Gamma_{0}$
which corresponds to a magnetic field of $B = 100$ Gauss. Furthermore, we take equal doping density for both nuclear sites 
such that $\eta^{(1)}_{13} = \eta^{(2)}_{13}$. 

The red solid curve of Fig.~\ref{fig7} shows the NFS time spectrum in this case for equal doping density of the 
two nuclear sites. The spectral response is found to be distinctly different from the homogeneous case due to an 
interplay of two set of quantum interferences---intra-nuclei interference due to the Autler-Townes splitting of the resonances with formation of 
dressed states (\ref{eq8}) for each group of the nuclei and inter-nuclei one due to the two different sets of nuclei with different hyperfine splittings. 
As $\delta_{i}(B) \sim \Omega_{c}$ the intra-nuclei quantum interference in the two groups of nuclei with and without $\delta_{i}(B)$ is of the order 
of $\sqrt{2}\Omega_{c}$ and $\Omega_{c}$ respectively. In Fig.~\ref{fig7} we find the initial beating pattern as a result of interference 
between two groups of nuclei with different frequencies which however is smeared out gradually with time. Furthermore, contrary to 
the uniform doping case, at longer times the behavior of the NFS time spectrum is no more governed by the decoherence rate of the probe transition. 
This feature can be attributed to loss of coherence among scattered field amplitudes from the two groups of nuclei with different detunings. 
The shallower slope in this case thus suggests that the system encounters an additional inhomogeneous broadening owing to the 
two different doping sites. Note that such distinct behavior of NFS with two-site doping arise only when $\delta_{i}(B) \geq \Omega_{c}$.
In the other limit, the Autler-Townes splitting dominates and the nuclei in both sites  would experience  a beating $\sim \Omega_{c}$ and would become indistinguishable thereby 
resulting in a NFS time spectrum similar to that of a uniformly doped sample.  

\subsection*{Pulsed control field}

\begin{figure}[!h]
\scalebox{0.8}{\includegraphics{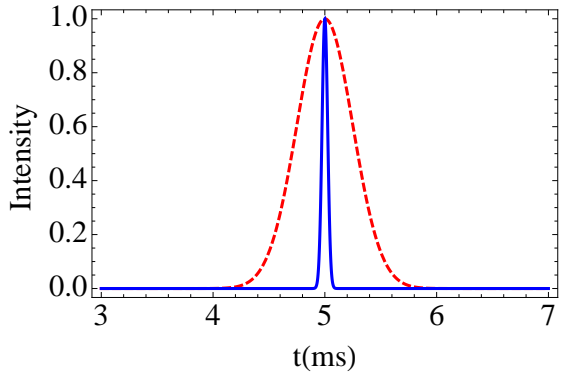}}
  \caption{\label{fig8} (Color online) Time domain pulse shape of 
  the control and probe fields shown by the dashed (red) and solid (blue) line. The control pulse has $10$ times the width 
  of  the probe pulse with $\Omega_{c0} = 10^{8}\Gamma_0$ and $t_{c0} = 5$ ms, $\tau_{c} = 0.5$ ms, 
  while the probe parameters are $\Omega_{p0} = 10^{6}\Gamma_0$ and $t_{p0} = 5$ ms, $\tau_{p} = 0.05$ ms, respectively. }
\end{figure}

In the above discussion  we have assumed 
that the control field is a continuous wave. However, a cw laser source in the VUV region is currently 
available only within limitations. The KBe$_2$BO$_3$F$_2$ crystals \cite{chen09} have been successful in 
generating narrow-band VUV radiation via harmonic generation owing to their wide transparency and 
large birefringence necessary for phase-matched frequency conversion processes in this frequency 
region \cite{Tog03,Nom90}. A quasi-cw coupling VUV laser at around $160$-nm wavelength could also be generated by 
the sum frequency mixing in metal vapors or driving a KBe$_2$BO$_3$F$_2$ crystal with a 
Ti: sapphire laser \cite{Tog03,Nolt90}. To circumvent these limitations we propose to use instead of cw a 
VUV pulse of width much broader than the VUV probe pulse with complete overlap of the two in 
time domain as shown in Fig. \ref{fig8}.

Both control and probe Rabi frequencies in the
Bloch equations (\ref{eq3a}) are in this case space and time dependent.
Hence, the field equations for propagation (\ref{eq3b}) have to be now solved for both probe and control with additional 
initial and boundary conditions for the control field given by
\begin{equation}
\label{eq10a}
\Omega_{c}[0,t] = \Omega_{c0}\exp\left[-\left(\frac{t-t_{c0}}{\tau_{c}}\right)^2\right] , \ \ \Omega_{c}[z,0] = 0.
\end{equation}
The NFS signal is measured solely from the scattering of the probe pulse with the control pulse 
influencing the dynamics during its presence by formation of a time-dependent dressed state. 
In principle, by varying the overlap time of the probe and 
control pulses one can achieve a stimulated Raman adiabatic passage (STIRAP)
in this configuration. This has already been proposed for such configuration of nuclear levels 
involving keV- and MeV-transition energies in Refs.~\cite{Wen_PL11,WenPRC13}.  
\begin{figure}[!h]
\scalebox{0.8}{\includegraphics{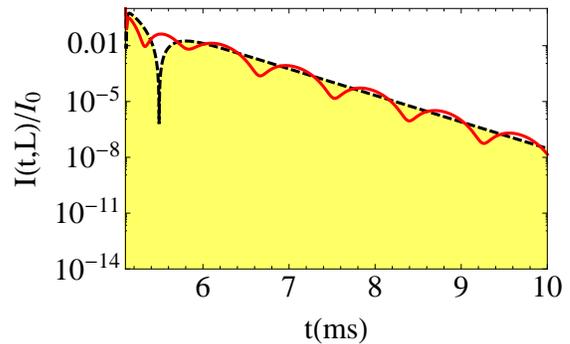}}
  \caption{\label{fig9} (Color online)  The time domain spectrum of NFS for three level configuration with a two pulsed-field setup. 
  The black dotted with yellow shading (red solid) curve  is for a single-doping (two-doping)  site case.
  Due to doping at a different site an additional shift of $\delta_{2}(B) = 10^{8}\Gamma_{0}$ of the energy level is induced 
  in the second group of nuclei. Here we have considered equal doping density for the two groups such that 
  $\eta^{(1)}_{13} = \eta^{(2)}_{13} = \eta_{13}/2 \simeq  227$~Hz/cm 
  and $\eta^{(1)}_{12} = \eta^{(2)}_{12} = \eta_{12}/2  \simeq  22.7$~Hz/cm 
 for $\xi \simeq 10^6$. The spontaneous emission rates are same as that of Fig. ~\ref{fig7}
 The control and probe fields are equally detuned from the level $|3/2,3/2\rangle$ by $\Delta = 10^{5}\Gamma_0$. }
\end{figure}

In Fig.~\ref{fig9} the black dotted curve shows the time spectrum of NFS from $^{229}$Th doped VUV crystal for 
one doping site with both control and probe as pulsed fields. During the time when the 
control field is present the behavior of the NFS spectra is quite similar to that seen in Fig.~\ref{fig7} and we see
formation of quantum beat. However, once the control and probe no longer overlap at $t\sim 6$ ms, the spectra become 
similar to that of two level nuclei without any sensitivity to the probe detuning as discussed earlier. This time-dependent behavior of the beating is 
related to the time dependence of the dressed frequency $\tilde{\Omega}(t)$ \cite{Shore, Das_PRA11} and of the dressed states (\ref{eq8}).
The time-dependent splitting of the resonances  creates two transitions for the probe to interact with
for the duration of the control pulse and determines the emission characteristics. As for the long time behavior 
of the NFS spectra, it is found to be similar to the two level case, with the response being dominated by the 
decoherence rate of the probe transition.

For the case of two doping sites in the VUV crystal, 
the space- and  time dependent propagation of the control field have now contribution from coherences 
of both groups. Thus the field equations in this case become, 
\begin{eqnarray}
\label{eq10}
\partial_{z}\Omega_{p} & +& \frac{1}{c}\partial_{t}\Omega_{p} =  i a_{13}(\eta^{(1)}_{13}\rho^{(1)}_{13}+\eta^{(2)}_{13}\rho^{(2)}_{13}),\nonumber\\
\partial_{z}\Omega_{c} & +& \frac{1}{c}\partial_{t}\Omega_{c} =  i a_{12}(\eta^{(1)}_{12}\rho^{(1)}_{12}+\eta^{(2)}_{12}\rho^{(2)}_{12}),
\end{eqnarray}
with the Bloch equations given by (\ref{eq3b}), where the Hamiltonian now has an additional intrinsic magnetic field 
dependent detuning $\delta_{i}(B) \ (i = 1,2)$ for the second group of nuclei as was discussed at length for the case 
of a cw control field. For simulation we consider equal doping density for the two groups such that 
$\eta^{(1)}_{12} = \eta^{(2)}_{12}$ and $\eta^{(1)}_{13} = \eta^{(2)}_{13}$. 

The NFS time spectrum is obtained from the field equation for the probe as shown by 
the solid (red) curve in Fig. \ref{fig9}. For a short  time when the probe and control pulses 
overlap, we find the behavior similar to that of Fig. \ref{fig7} governed by the interplay of two quantum 
interferences. The intra-nuclei interference arises from the two transitions created by time-dependent dressed states, 
while the inter-nuclei interference comes into play due to different hyperfine splitting in two groups of $^{229}$Th nuclei occupying 
the two different nuclear sites. However, once the control field and the probe cease to overlap, the dressed states and 
Autler-Townes doublet vanishes. Thus there is no further intra-nuclei interference and the NFS probe response bear 
signature of only inter-nuclei interference among the two groups of nuclei with different hyperfine splitting owing 
to different doping sites. Note that this behavior is similar to the two level case, as in absence of the 
control field the probe simply interacts with a single transition in the nuclei. 
Furthermore, the envelope of the beating pattern is found to follow the decoherence rate 
of the probe transition in analogy to the simple two level case. 

\subsection{Nuclear ensemble in a three level $V$-configuration}

\begin{figure}[h]
\scalebox{0.4}{\includegraphics{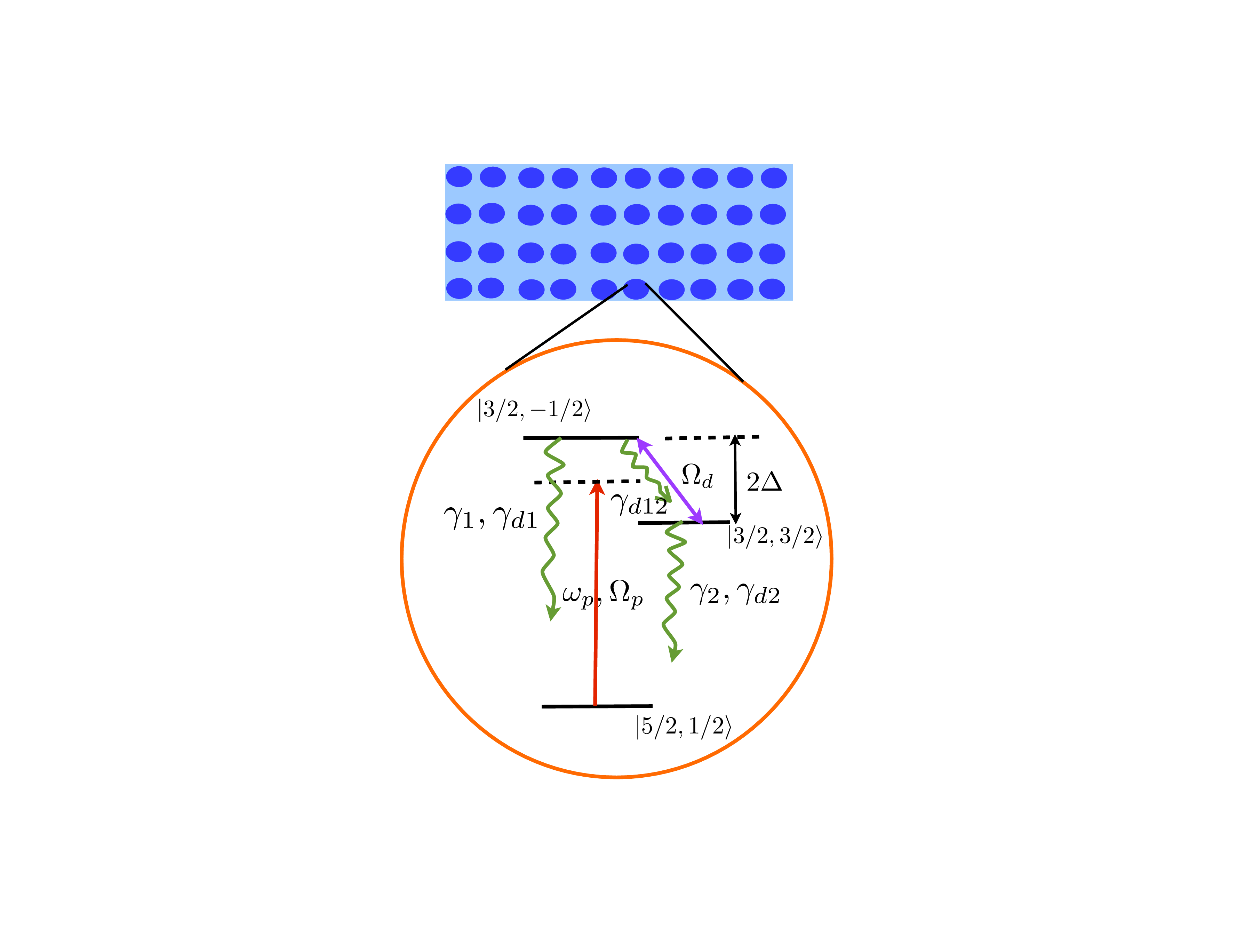}}
  \caption{\label{fig10} (Color online) VUV excitation of the doping thorium nuclei in a three level V-configuration. 
  The pulse with frequency $\omega_{p}$ is equally detuned from the upper doublet by $\Delta$. The time-dependent 
  driving pulse $\Omega_{d}(t)$  creates a time-dependent coherent superposition of the upper levels. 
  Population trapping in the excited doublet occurs as long as drive and probe pulse overlap. See text for further explanations.}  
\end{figure}

We next consider an ensemble of doping nuclei in the VUV-transparent crystal driven by a strong VUV probe pulse  that 
 couples two nearly degenerate upper levels to a common ground level
forming a $V$-configuration as shown in Fig. \ref{fig10}. Such a $V$-level scheme 
has been extensively investigated in atomic quantum optics. For instance, it was shown that under the restrictive 
condition of non-orthogonality of the dipole transitions, decay-induced coherence among the upper levels is generated 
in such a scheme \cite{agarwal, Kiff_10}. This coherence may lead to  fascinating phenomena like 
lasing without inversion \cite{LWI1}, breaking of detailed balance \cite{Scully10} and manipulation of resonance profiles \cite{Das_PRAL11}.
However, in our system the magnetic dipole transitions from the two upper to a common 
lower nuclear state are left and right circularly polarized and therefore orthogonal. To drive 
both the transitions with a common field it is thus important to select the suitable polarization of the VUV pulse,
which in this case is chosen to be along the $\hat{x}$ direction. 
Here we are interested in achieving population trapping among the upper levels by creating initial 
coherence among them via an additional VUV pulse as depicted in Fig. \ref{fig10}. 
In the following we denote this additional field as  the driving field. 
The interaction Hamiltonian in the dipole approximation and in a frame rotating with the frequency $\omega_{p}$ 
of the probe for the $V$- scheme has a form similar to Eq. (\ref{eq1}) with $j = 1, 2$ and the interaction 
part modified to
\begin{eqnarray}
\label{eq11}
\hbar\sum_{j}(\frac{\Omega_{j}}{2}(\vec{\varepsilon}_{fj}\cdot\vec{\varepsilon}_{nj})S^{+}_{j}+H.c.) & \rightarrow & \hbar\frac{\Omega_{1}[z,t]}{2}[(\hat{x}\cdot\vec{\sigma}_{-})S^{+}_{1}\nonumber\\
+ (\hat{x}\cdot\vec{\sigma}_{+})S^{+}_{2}] & - & \hbar\frac{\Omega_{d}}{2}[S^{+}_{1}S^{-}_{2}]+H.c.,\nonumber\\
\end{eqnarray}
where $\vec{\sigma}_{\pm} = (\hat{x}\pm i\hat{y})$, $\Omega_{p}$ and $\Omega_{d}$ are the Rabi frequencies of the probe and driving field, respectively. 
The detunings in the Hamiltonian (\ref{eq1}) are now $\Delta_{1} = \omega_{13}-\omega_{p}$ and $\Delta_{2} = \omega_{p}-\omega_{23}$ 
and the nuclear raising (lowering) operators are defined  in 
the three level basis $\{|3/2,-1/2\rangle = |1\rangle, |3/2,3/2\rangle = |2\rangle,|5/2,1/2\rangle = |3\rangle\}$
as $S^{+}_{1} = |1\rangle\langle 3|,  S^{+}_{2} = |2\rangle\langle 3|, S^{-}_{k} = (S^{+}_{k})^\dagger$, with $k = 1,2$. 
The population relaxation in the $V$-level configuration is given by the Louvillian operator (\ref{eq2}) with
$\gamma_{1}$ and $\gamma_{2}$ being the population relaxation rates of the nuclear levels 
$|3/2,-1/2\rangle$ and $|3/2,3/2\rangle$ to the common ground levels 
$|5/2,1/2\rangle$, respectively. The decoherence of the relevant transitions is incorporated in 
the dynamics phenomenologically by means of the decoherence matrix
\begin{equation}
\label{eq13}
\mathcal{L}_{d}\rho = -\left[ 
\begin{array}{cccc}
  0 & \gamma_{d12}\rho_{12} & \gamma_{d1}\rho_{13}\\
\gamma_{d12}\rho_{21} & 0 &\gamma_{d2}\rho_{23}\\\
  \gamma_{d1}\rho_{31} & \gamma_{d2}\rho_{32}& 0
\end{array}  
\right],
\end{equation}
where $\gamma_{d12}$, $\gamma_{d1}$ and $\gamma_{d2}$ are the decoherence rates of the transitions 
$|1\rangle \rightarrow |2\rangle$, $|1\rangle \rightarrow |3\rangle$ and $|2\rangle \rightarrow |3\rangle$, respectively.

\begin{figure}[!h]
\scalebox{0.8}{\includegraphics{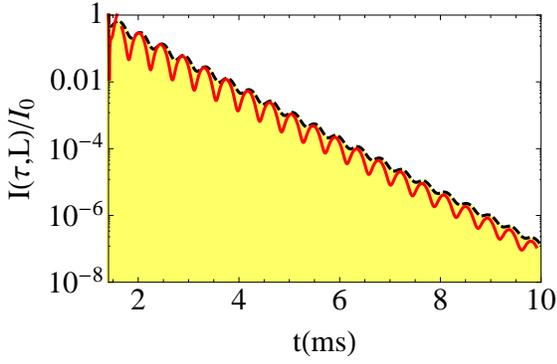}}
  \caption{\label{fig11} (color online) The time domain spectrum of NFS for three level $V$-configuration in a two-field setup. 
  The black dotted (red solid) curve is in absence (presence) of the external field coupling between the upper doublet.  
  The parameters used for computation are $\gamma_{1} = 0.267\Gamma_{0}$, $\gamma_{2} = 0.0889\Gamma_{0}$, 
  $\eta_{13} \simeq  22.7$~Hz/cm, $\eta_{23} \simeq 68$~Hz/cm for 
  $\xi \simeq 10^{6}$, $\Omega_{d0} = 10^{8}\Gamma_{0}$, $t_{d0} = 1$ ms, $\tau_{d} = 0.1$ ms, 
  $\Omega_{p0} = 10^{9}\Gamma_{0}$, $t_{p0} = 1.1$ ms and $\tau_{p} = 0.1$ ms.}
\end{figure}

To obtain the NFS time spectra from the nuclear ensemble in such a $V$-level 
scheme we numerically compute the output field from the Maxwell-Bloch equations 
(\ref{eq3b}) and (\ref{eq3c}) without summation over the index $\alpha$ (we consider 
a single-doping site scenario for this scheme) and with $j = p,  \ l = 1,2 , \ k = 3$. The field propagation
equation becomes 
\begin{equation}
\label{eq14}
\partial_{z}\Omega_{p}  +  \frac{1}{c}\partial_{t}\Omega_{p} =  i( \eta_{13}a_{13}\rho_{13}+\eta_{23}a_{23}\rho_{23}),
\end{equation}
with contributions from coherences along both the transitions. The NFS time spectrum generated 
by numerical computation of (\ref{eq14}) and the Bloch equations are shown in Fig. \ref{fig11}.   
Here we have assumed the probe field given by (\ref{eq4}) to be positively and negatively detuned 
from the level $|3/2,3/2\rangle$ and $|3/2,-1/2\rangle$ respectively, by $\Delta = 10^{8}\Gamma_{0}$
and that all the population is initially in the ground state $\rho_{33} (0) = 1$. The decoherence
rates are taken as $\gamma_{d12} = 2\pi\times 45$ Hz, $\gamma_{d1} = 2\pi\times 158$ Hz and 
$\gamma_{d2} =  2\pi\times 142$ Hz in accordance with the estimates for $^{229}$Th in CaF$_2$ \cite{Kaz12}. 
For numerical purposes we have assumed that the driving field which couples the upper levels 
has a Gaussian shape 
\begin{equation}
\label{15}
\Omega_{d}[t] = \Omega_{d0}\exp\left[-\left(\frac{t-t_{d0}}{\tau_{d}}\right)^2\right]. 
\end{equation}

The dotted black curve in  Fig. \ref{fig11} shows the behavior of the NFS time spectrum
in absence of any coupling of the upper levels and after passage of the excitation pulse. 
We find quantum beat features owing to intra-nuclei quantum interference 
among the two transitions of the $V$-configuration. The beats are of the order 
of $2\Delta$ with an envelope governed by the highest decoherence rate 
in the system. 
The red solid curve in Fig. \ref{fig11} depicts the behavior of NFS time spectra, when the upper levels are 
coupled by the driving field $\Omega_{d}$ which overlaps with the probe for a short period of time ($\sim 0.1$ ms). 
The quantum beat pattern that we find in this case is similar to that found in absence of the driving 
field, however now with enhanced amplitudes. Thus presence of coupling among the upper doublet 
does not show any significant influence on the probe NFS spectrum. This can be understood as follows.
The probe field overlaps with the driving field for a very short time during its propagation through the 
medium and thus any influence on the NFS spectrum will be transient and may be visible at initial 
moments of the scattering. Furthermore, the decoherence of $\rho_{12}$ spoils the coherence effects
within ms of the its creation thereby erasing any signature of possible interplay of the two fields in the 
NFS spectra. Finally, considering the drive field to be strong enough to form a dressed state of the 
upper doublet like that discussed earlier for the $\Lambda$ configuration, for parameters used in the figure 
the dressed frequency would be on the same order as the probe detuning $\Delta$. Hence,  no additional beating induced for the time of overlap would be observed. 

\begin{figure}[!t]
\centering
\mbox{\subfigure{\includegraphics[width=4.3cm]{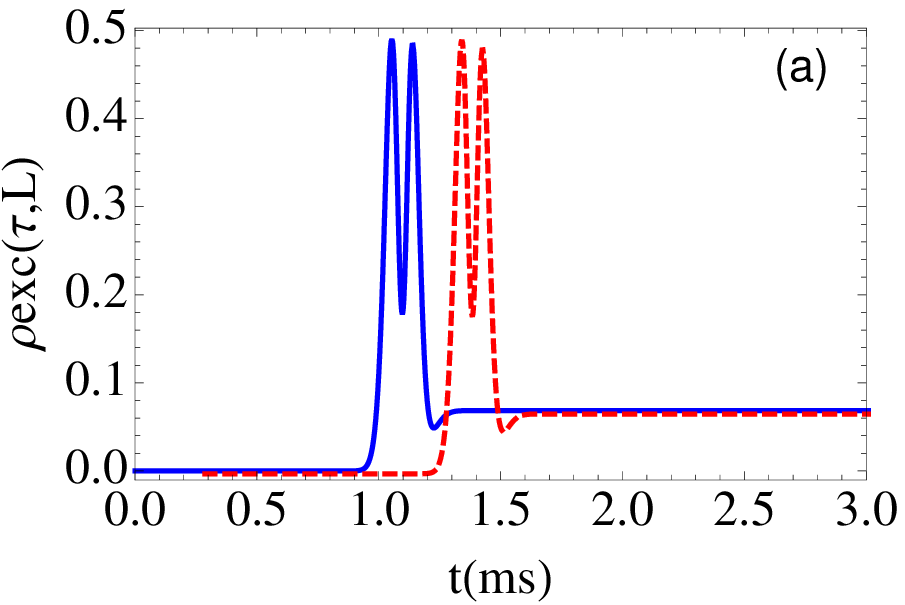}}
\hspace{0.03cm}
\subfigure{\includegraphics[width=4.3cm]{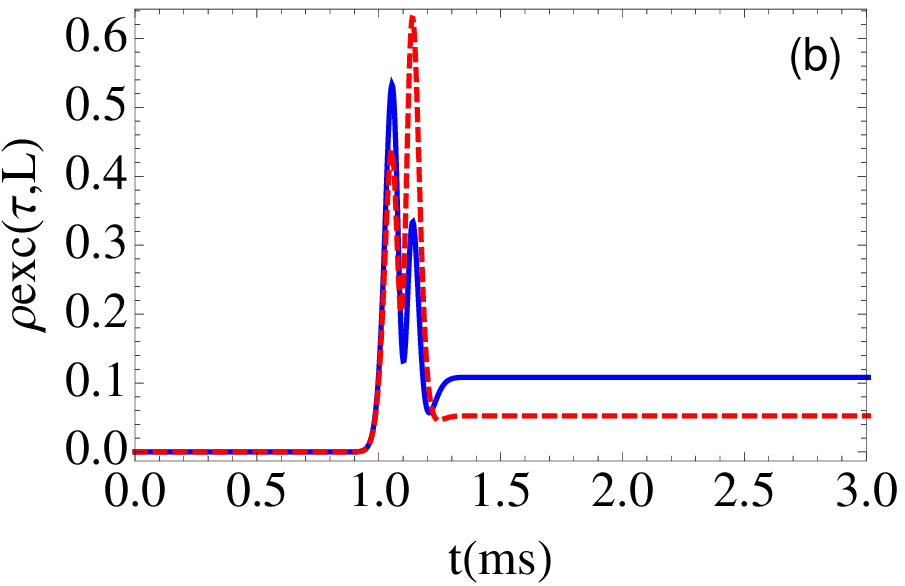}}}\\
\mbox{\subfigure{\includegraphics[width=4.3cm]{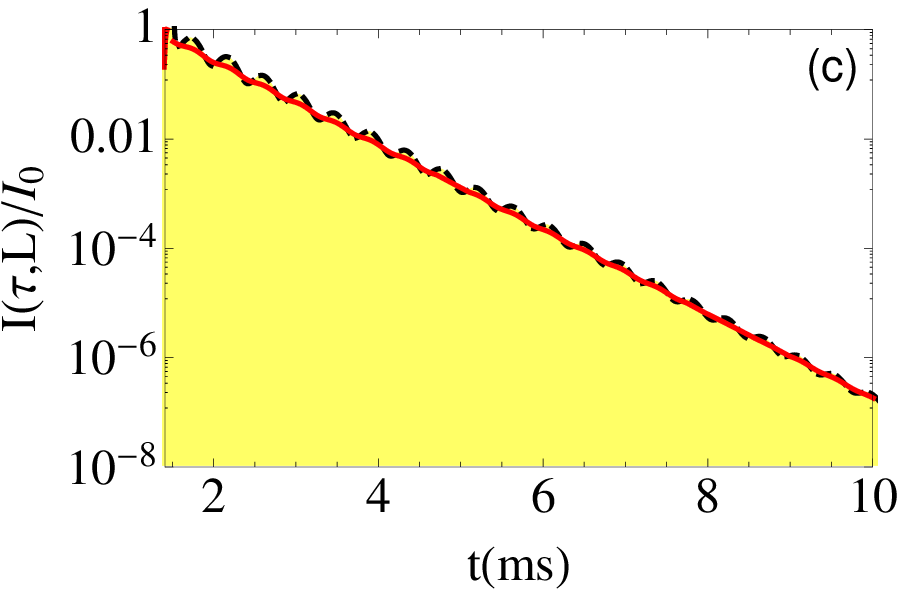}}
\hspace{0.03cm}
\subfigure{\includegraphics[width=4.3cm]{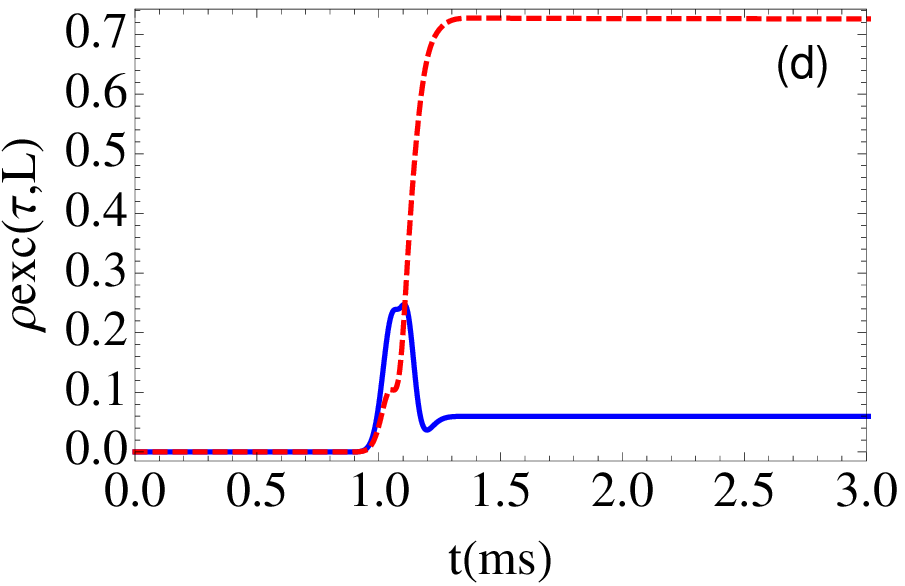}}}
  \caption{\label{fig12} (color online) Population dynamics of the upper state doublets in $V$-configuration in (a) absence 
  and (b, d) presence of coherent coupling among the upper states.  
  The solid (blue) and dashed (red) curves illustrate  $\rho_{11}$ and $\rho_{22}$. 
  The amplitude of the driving field is different in 
  (b) and (d) being respectively, $\Omega_{d} = 10^{8}\Gamma_{0}$ and $\Omega_{d} = 10^{9}\Gamma_{0}$.
  All other parameters are the same as in Fig.~\ref{fig11}.
  In (a) the dashed (red) line is shifted along the time axis by 0.5 ms for better visibility. 
  (c) shows the NFS spectra corresponding to the population dynamics in (d).}
\end{figure}

Although we do not see any dominant signature of the coherence among the upper doublet  
in the NFS spectra, a study of the population dynamics of the upper states shows that population trapping is achieved.
 In Fig. \ref{fig12} 
we plot the populations of the upper states
$\rho_{11}$ and $\rho_{22}$ in presence (b,d) and absence (a) of the driving field. Without driving field 
among the states $|1\rangle$ and $|2\rangle$ we find in Fig. \ref{fig12} (a) that for a strong excitation 
probe ($\Omega_{p0} = 10^{9}\Gamma_{0} = 0.1$ MHz) the populations undergo Rabi oscillations 
between the states $|1\rangle \rightarrow |3\rangle$ and $|2\rangle \rightarrow |3\rangle$ during 
the pulse duration. The population dynamics is seen to be exactly the same for the two upper states.  
When the excitation pulse is gone, the ground state $|3\rangle$ remains mainly populated with 
about $5\%$ of population in the upper states. The flat tail arises because these remaining 
population decays with the isomeric half-life that is much longer than our (ms) timescale. 
However, this behavior changes significantly in presence of the driving field coupling the states 
$|1\rangle$ and $|2\rangle$ and the population dynamics becomes dependent on the relative 
strength of the drive and the probe. 

For a weaker driving field compared to the probe $\Omega_{d} = 0.1\Omega_{p0}$, we see in Fig. \ref{fig12} (b) 
that the population dynamics of upper doublet is no more symmetric for the probe pulse duration. 
The drive creates a coupling between the states $|1\rangle$ and $|2\rangle$ via the coherence $\rho_{12}$
which leads to population exchange among the states once any one of them is populated. The time-dependent dynamics then shows pronounced asymmetry in the Rabi oscillations of the two states as can be seen in the figure. 
The generated coherence is strongly influenced  by the decoherence of the $|1\rangle\leftrightarrow|2\rangle$ transition and 
depends on the Rabi frequency of the driving field. In the present case it is weak due to a weaker Rabi coupling and is short lived 
due to the decoherence effect. Thus at later times, when the probe and drive cease to overlap, we find 
only $10\%$ and $5\%$ of the initial population still remaining in the upper states  $|1\rangle$ and $|2\rangle$, respectively. 
The remaining population then decays with the isomeric half-life
($\sim$ hrs) that is much longer than our (ms) timescale. Thus effectively we have moderately enhanced 
the trapped population in $|1\rangle$ due to the initial coherence among the doublet upper states.

The behavior of the upper states populations  dramatically changes however  
when the driving field has a strength of the order of the probe field. 
The created coherence is stronger and leads to strong coupling 
among the upper doublet even though short lived due to decoherence.   
Figs.~\ref{fig12}(c) and  \ref{fig12}(d) illustrate the NFS spectra and the population dynamics, respectively.
From Fig. \ref{fig12}(d) we see that the population of $\rho_{11}$ increases gradually during the overlap of the probe and drive fields but then 
start decreasing as the fields separate. Here $\rho_{11}$ essentially shows a Lorentzian 
absorption peak in presence of both the fields with small amount of population left in their absence.
The population that goes in $\rho_{11}$ due to strong coupling to $|2\rangle$ gets quickly distributed between the two upper 
states. In addition, state $|2\rangle$ becomes more and more populated as $\rho_{11}$ decreases due to decoherence of $\rho_{12}$.  
As such, while $\rho_{22}$ in presence of the drive and probe overlap increases gradually and attains a value similar to $\rho_{11}$ 
when the fields gradually separate, it does not decrease and rather accumulates further population which eventually 
get trapped due to long isomeric lifetime.  We see from Fig. \ref{fig12}(d) that almost $70\%$ of the total population gets trapped 
in  state $|2\rangle$. The obtained trapping persists for a long time as seen by the flat tail of the 
population since $|2\rangle$ decays with isomeric half-life which is much longer than the 
(ms) timescale considered for our NFS calculations. 
\begin{figure}[!h]
\centering
\mbox{\subfigure{\includegraphics[width=4.3cm]{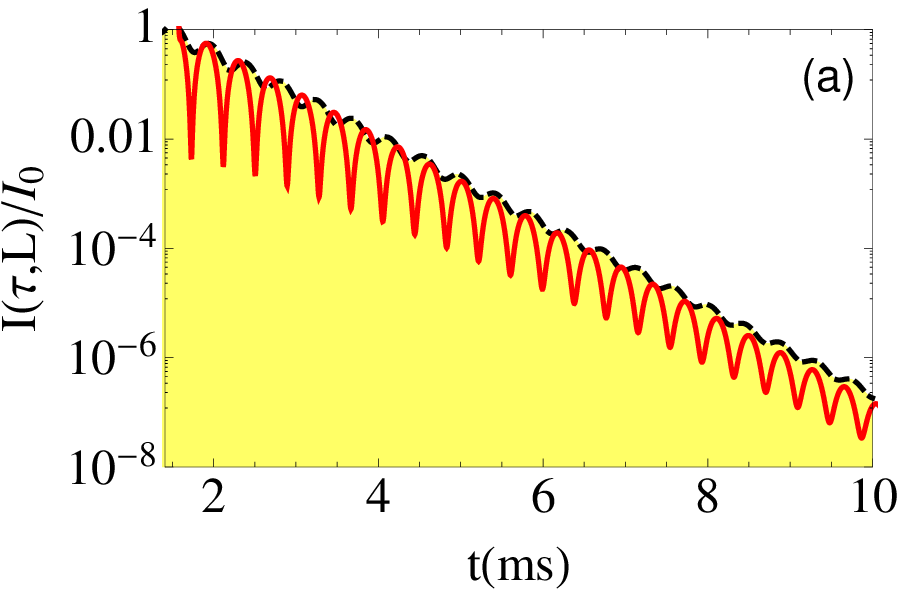}}
\hspace{0.03cm}
\subfigure{\includegraphics[width=4.3cm]{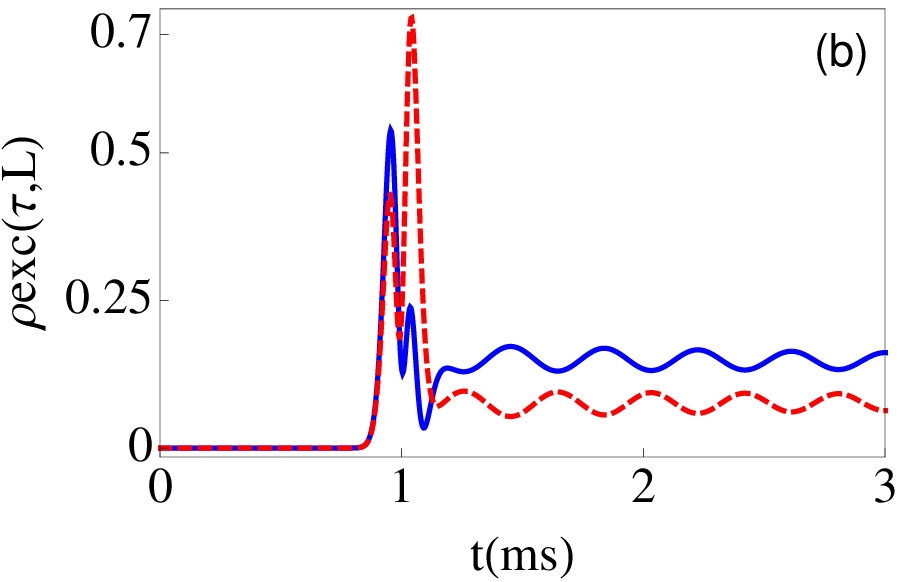}}}
  \caption{\label{fig12a} (color online) (a) NFS time spectra in absence (dashed black line with shading)
  and presence (solid red line) of an external DC field coupling of the upper level doublet.
 (b) The solid (blue) and dashed (red) curves illustrate  $\rho_{11}$ and $\rho_{22}$
  when the upper states are coupled by a DC field with Rabi frequency $\Omega_{d} = 10^{8}\Gamma_{0}$.
  All other parameters are considered to be same as for Fig.~\ref{fig11}.}
\end{figure}

In Fig. \ref{fig12a} we plot the NFS time spectrum and the population dynamics for the $V$-configuration with a DC 
drive field coupling the upper level doublets. The behavior of the NFS spectra as seen from the red solid curve 
in Fig. \ref{fig12a} (a) is similar to that of Fig. \ref{fig11} for a pulsed driving field. In the DC field case the only 
difference is a  larger beating amplitude. The similarity arises due to the fact that the 
dressing of the upper doublets with the DC field gives for the considered parameters a dressed frequency of the same order of $\Delta$ and thus 
no additional feature appear in the NFS spectra. If however, the Rabi frequency of the DC field would be such that the dressed 
frequency is $>>\Delta$, then the beating frequency were given by the dressed frequency in the NFS spectra. The population dynamics of the 
upper state doublet in the presence of the DC field coupling is shown in Fig. \ref{fig12a}(b). We find the behavior of the 
population to be similar to the case when the doublet was coupled by a pulse driving field for the 
probe duration, see Fig. \ref{fig12}(b). However, for longer times we find population oscillation of the state $|1\rangle$ and $|2\rangle$
otherwise absent in the pulsed field coupling case. Furthermore, we find almost $15\%$ and $10\%$ of initial population 
trapped in the $|1\rangle$ and $|2\rangle$ states. 

In practice, one can achieve population trapping  in the overlap region of the two laser focal spots. Considering experimentally realizable spot sizes for VUV lasers \cite{Rel10}, we expect 
about $10^{11}$ nuclei to be addressed for a $0.1$ cm ensemble depth. Thus 
almost $10^{10}$ nuclei can be trapped  in state $|2\rangle$ and decay with the natural lifetime of the isomeric state despite the NFS setup. This in principle
can be harnessed towards creating controlled subradiance in such nuclear isomers. Note that a similar subradiance 
phenomenon has been recently shown in dilute cloud of cold atoms \cite{Kaiser}. Thus further studies in this
direction  contribute to the new field of collectivity-induced quantum optical phenomena in nuclear isomers. 

\subsection{Nuclear ensemble in a four level configuration}
In the case of $^{229}$Th, the nuclear level structure will have more than two or three specific 
levels owing to the hyperfine interactions with the possibility of multiple transitions sharing the same polarization.  Additionally, the 
spacing between hyperfine levels for a particular angular momentum $I$ can be such 
that they cannot be resolved with the current bandwidth of lasers. All this conditions warrant for the 
study of NFS from nuclear ensemble with multiple transition which are either degenerate 
or near degenerate. Such studies has been already done extensively in $^{57}$Fe---which is also a favorable 
test bed of several other studies related to NFS in nuclei. 
Here we investigate the NFS time spectra of $^{229}$Th in similar spirit with the objective to find 
signatures of coherence and quantum interference present due to multiple transition pathways. 
\begin{figure}[!h]
\scalebox{0.25}{\includegraphics{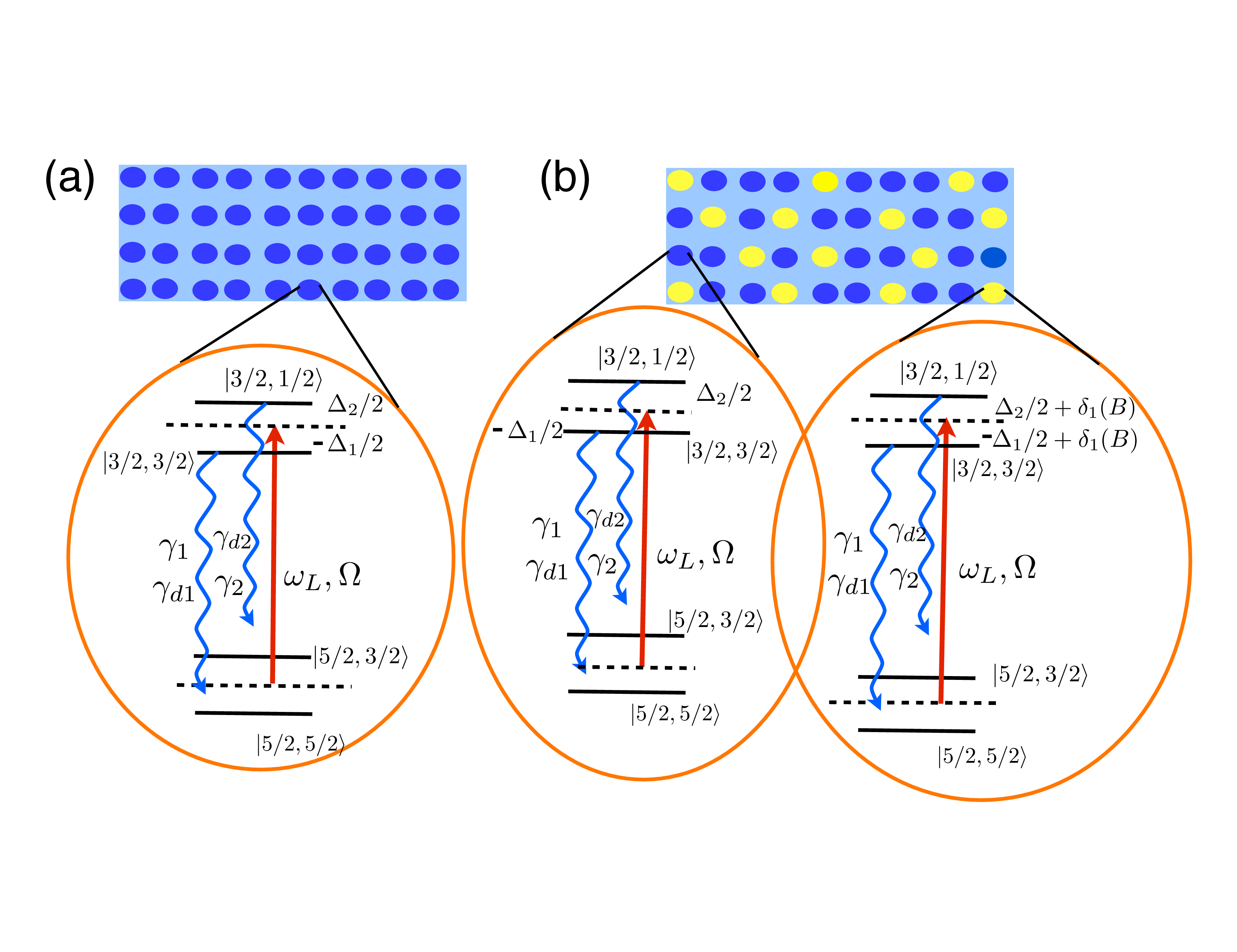}}
  \caption{\label{fig4} (Color online) (a) VUV excitation of the doped $^{229}$Th in the crystal lattice in the case of a single doping site. The laser 
  with frequency $\omega_{c}$ is tuned near the $|5/2,5/2\rangle \rightarrow |3/2,3/2\rangle$ and  
  $|5/2,3/2\rangle \rightarrow |3/2,1/2\rangle$ transitions. The relaxation and decoherence rates $\gamma_{1}$, $\gamma_{d1}$ $\gamma_{2}$ and $\gamma_{d2}$ 
  are assumed for the addressed transitions. 
  (b) The case of two doping sites. The blue group of thorium nuclei is influenced by a different environment than 
  the yellow one which experience the additional detunings $\delta_{1}$ and $\delta_{2}$ 
  as a function of the magnetic field. See text for further explanations.}
\end{figure}
We consider in the following a  basic model of four 
level nuclei with two upper and two lower nuclear levels as shown in Fig. \ref {fig4}(a). 
The nuclear levels are chosen such that given the selection rules and polarization of the incident VUV pulse  
only two parallel and near degenerate magnetic dipole transitions $|1\rangle \rightarrow |3\rangle$ and 
$|2\rangle \rightarrow |4\rangle$ are driven. 
For $^{229}$Th such a four level model can be formed by the states $\{|3/2,3/2\rangle, |3/2,1/2\rangle, |5/2,5/2\rangle, |5/2,3/2\rangle\} =  \{1,2,3,4\}$
driven by a left circularly polarized VUV laser pulse. Since we for now consider only one doping site 
in the VUV crystal we drop the index $\alpha$ on the density operator in the further discussion. 
  
The interaction Hamiltonian for our four level model in the dipole approximation and in a frame rotating with 
the probe laser frequency $\omega_{p}$ is given by (\ref{eq1}) with $j = 1,2$ for the two possible transitions. 
The nuclear raising (lowering) operators are now defined in a four level basis as 
$S^{+}_{1} = |1\rangle\langle 3|$, $S^{+}_{2} = |2\rangle\langle 4|$, $S^{-}_{k} = (S^{+}_{k})^{\dagger}$,  $(k = 1,2)$. 
The laser detunings in the Hamiltonian are respectively 
$\Delta_1 = \omega_{13}-\omega_{p},  \Delta_2 = \omega_{24}-\omega_{p}$ and the Rabi frequency which is assumed 
to be same for both  transitions, $\Omega_1 = \Omega_2 = \Omega_p$. 
The population relaxation of the hyperfine levels of the isomeric state is included in the dynamics by
the  Louivillian operator (\ref{eq2}) with $j = 1,2$,
where now $\gamma_1$ and $\gamma_2$ are the decay rates of the levels 
$|1\rangle$ and $|2\rangle$ to  $|3\rangle$ and $|4\rangle$, 
respectively. The effect of environmental decoherence is incorporated in 
the dynamics of the model phenomenologically via the decoherence matrix
\begin{equation}
\label{eq18}
\mathcal{L}_{d}\rho = -\left[ 
\begin{array}{ccccc}
  0 & 0 & \gamma_{d1}\rho_{13} & 0\\
0& 0 & 0 & \gamma_{d2}\rho_{24}\\
\gamma_{d1}\rho_{31} & 0 & 0 & 0\\
\gamma_{d2}\rho_{42} & 0 & 0 & 0\\
\end{array}  
\right],
\end{equation}
where $\gamma_{d1}$ and $\gamma_{d2}$ are the decoherence rates of the transitions $|1\rangle \rightarrow |3\rangle$ 
and $|2\rangle \rightarrow |4\rangle$, respectively.

To obtain the time spectra of the NFS signal from the nuclear ensemble in such a $4$-level scheme we 
numerically evaluate the output field from the Maxwell-Bloch equations (\ref{eq3a}) and (\ref{eq3b})
where now the field equations (\ref{eq3b}) contain contributions from coherences along both  transitions, 
 \begin{eqnarray}
 \label{eq19}
\partial_{z}\Omega_{1} +  \frac{1}{c}\partial_{t}\Omega_{1} =  i(a_{13} \eta_{13}\rho_{13}+a_{24} \eta_{24} \rho_{24}).
\end{eqnarray}
We numerically solve Eq. (\ref{eq19}) with the initial condition that the hyperfine levels of the lower 
state are equally populated, i.e., $\rho_{33}(0,z) = 0.5, \rho_{44}(0,z) = 0.5$ and with the initial and boundary conditions 
on the probe field given by Eq. (\ref{eq4}). We consider the decoherence rates of $\gamma_{d1} = 2\pi\times 84$ Hz 
and $\gamma_{d2} = 2\pi\times 251$ Hz which have been estimated for these transitions in $^{229}$Th:CaF$_2$ \cite{Kaz12}.

\begin{figure}[!t]
\centering
\includegraphics[width=7cm]{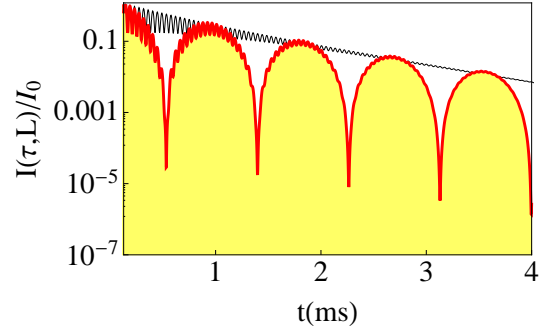}
  \caption{\label{fig5} (Color online) NFS time domain spectra  for a multilevel sample with near degenerate transitions. 
  The thin (black) and thick (red) curves with yellow shading are 
  for sample with one and two doping sites, respectively with laser detuning of $\Delta = 10^{9}\Gamma_{0}$.
  A magnetic field dependent detuning of  
  $\delta_{i}(B) = 10^{8}\Gamma_{0}$ in the case of two doping sites is considered.  
 Here, we have considered $\eta^{(1)}_{13} = \eta^{(2)}_{13} = \eta_{13}/2 \simeq  227$~Hz/cm and 
 $\eta^{(1)}_{24} = \eta^{(2)}_{24} = \eta_{24}/2 \simeq 137$~Hz/cm for 
  $\xi \simeq 10^{6}$. The spontaneous decay rates of the relevant transitions are respectively,
  $\gamma_{1} = 0.899\Gamma_{0}$ and $\gamma_{2} = 0.533\Gamma_{0}$.
  The probe pulse has a duration of  $\tau = 0.01$ ms with a Gaussian shape centered on $t_{0} = 0.1$ ms.}
\end{figure}

For an incident probe field of Rabi frequency $10^{6}\Gamma_{0}$ detuned to the transition 
$|2\rangle \rightarrow |4\rangle$ and $|1\rangle \rightarrow |3\rangle$ by $\Delta_{2} = -\Delta_{1} = 10^{9}\Gamma_{0}$ 
respectively, we obtain the scattered NFS intensity as shown by the thin (black) curve with shading (yellow) in Fig. \ref{fig5}. 
The signal shows quantum beat features with a frequency of the order of $10^{9}\Gamma_{0}$. The origin of this 
beating is attributed to the intra-nuclei quantum interference between the two transition pathways that the probe field interacts with. 
The envelope of the beating is seen to be governed by the largest of the two decoherence rates involved, i.e., by $e^{-2\gamma_{d2}t}$. 
Additionally, from the figure we see that the beating gets damped with time.  
%
\section*{Different doping sites}

For the case of two different doping sites we assume that the VUV crystal now comprises of two groups of the dopant nuclei
represented by the blue and yellow dots in Fig. \ref{fig4}. Thus the index $\alpha$ on the density operator in 
Eqs. (\ref{eq2}), (\ref{eq3a}) and (\ref{eq3b}) takes the values $1$ and $2$ for the two groups. The effect of environmental
perturbation on the coherences is included via the decoherence matrix (\ref{eq18}) for each group of nuclei.  
The nuclei of the second group have an additional shift $\delta_{k}(B)\  (k = 1, 2)$ of the isomeric level as shown in Fig. \ref{fig4}(b).  
 In the Hamiltonian description in Eq. (\ref{eq1}) of the four level system   the detunings will now be replaced by 
$\Delta_{1} \rightarrow \Delta_{1} + \delta_{1}(B)$ and $\Delta_{2} \rightarrow \Delta_{2} + \delta_{2}(B)$, respectively. 
As the probe VUV pulse will now simultaneously interrogate the nuclei in both crystal unit cell sites, the field equation will 
have contributions from the coherences related to both nuclear sites as well as from both the transitions of each group. 
The field equation thus involves several coherences in this case and takes the form
\begin{eqnarray}
\label{eq20}
\partial_{z}\Omega_{p} + \frac{1}{c}\partial_{t}\Omega_{p} & = &  i\bigg[ a_{13}(\eta^{(1)}_{13}\rho^{(1)}_{13}+\eta^{(2)}_{13}\rho^{(2)}_{13})\nonumber\\
&{}& +a_{24}(\eta^{(1)}_{24}\rho^{(1)}_{24}+\eta^{(2)}_{24}\rho^{(2)}_{24})\bigg],
\end{eqnarray}
where $\rho^{(1)}$ and $\rho^{(2)}$ are the density matrices for the first and second doping sites,  respectively. The dynamics of the 
coherences from each groups are governed by Eq. (\ref{eq3a}) with the Hamiltonian containing the modified detuning for the second group. 

As before, the probe pulse is assumed to be positively and negatively detuned to the transition $|2\rangle \rightarrow |4\rangle$ and 
$|1\rangle \rightarrow |3\rangle$ by $10^{9}\Gamma_{0}$ in the first group of nuclei. We furthermore assume that 
the nuclei in the second doping site have an additional hyperfine shift of the levels by $\delta_{k}(B) = 10^{8}\Gamma_{0}\  (k = 1,2)$ which corresponds 
to a magnetic field of about $100$ Gauss. Thus the nuclei in the second doping site have a laser detuning of $(10^{8}-10^{9})\Gamma_{0}$ for the
$|1\rangle \rightarrow |3\rangle$ transition and $(10^{9}+10^{8})\Gamma_{0}$ for the
$|2\rangle \rightarrow |4\rangle$ transition. The other parameters of the probe field are same as in the previously studied case. 
For computational purpose we consider equal doping of the 
two groups of nuclei such that $\eta^{(1)}_{13} = \eta^{(2)}_{13}$ and $\eta^{(1)}_{24} = \eta^{(2)}_{24}$.  

The NFS time spectrum evaluated numerically by means of Eq. (\ref{eq20})  
and the respective Bloch equations is shown by the solid (red) curve in Fig. \ref{fig5}. 
We find a signature of intra- and inter-nuclei quantum interference in the NFS signal.  
From Fig. \ref{fig5} we find that the intra-nuclei quantum interference which leads to beats of order $\Delta_{1,2}$, owing to
the two transition pathways in each nucleus, is modulated by inter-nuclei quantum interference 
of beats frequency $\sim\delta_{k}(B)$. The origin of inter-nuclei quantum interference
is the different hyperfine splitting of the two groups of nuclei. Similar to the uniform doping 
case here again we see that the envelope of the combined beating pattern is governed by the 
largest decoherence rate of the system.  

\section{Conclusion}
We have carried out an extensive theoretical study of NFS time spectra from an multi-level nuclear
ensemble of $^{229}$Th doped in a VUV crystal. Explicit results for three and four level configurations 
interacting with one and two optical fields were presented. In the three level case we have considered two different configurations namely the 
$\Lambda$ and $V$- level scheme interacting with two different VUV fields. We have shown that interference effects occurring in 
such configurations  offer  signatures of the isomer excitation advantageous for the more precise experimental determination of the transition 
energy. Our study shows that it is possible to coherently manipulate the signature of the quantum interference in the NFS signal 
by externally tuning the laser intensity and detuning.  Furthermore, the possibility of population trapping in the isomeric state has been investigated.
This can be utilized towards controlled subradiance generation in such nuclear isomers thereby opening new direction in quantum optical 
manipulation of collective phenomenon in nuclear systems.  Our study theoretically 
sustains the concept of nuclear coherent control and paves the way for further nuclear quantum optics applications with $^{229}$Th.

\section*{Acknowledgement}
The authors would like to thank W.-T. Liao for fruitful discussions. 


\end{document}